\documentclass[a4paper,11pt]{article}
\pdfoutput=1 

\usepackage{xcolor}
\usepackage{amssymb,amsmath}
\usepackage{mathrsfs}
\usepackage{amsfonts}
\usepackage{bbold}
\usepackage{graphicx,graphics}

\usepackage{fancyhdr}
\usepackage{lastpage}
\usepackage{layout}
\usepackage{longtable}
\usepackage{multirow}
\usepackage{slashed}
\usepackage{subfigure}

\usepackage[T1]{fontenc}
\usepackage{upquote}
\usepackage{cite,appendix}

\newcommand{\hp}{\mathbf{\hat p}}
\newcommand{\hk}{\mathbf{\hat k}}

\newcommand{\hr}{\mathbf{\hat r}}
\newcommand{\hn}{\mathbf{\hat n}}
\newcommand{\ha}{\mathbf{\hat a}}
\newcommand{\hb}{\mathbf{\hat b}}

\newcommand{\qqbar}{q{\bar q}}
\newcommand{\ttbar}{t{\bar t}}
\newcommand{\mttbar}{M_{t{\bar t}}}

\newcommand{\Sp}{{\mathbf S}_t}
\newcommand{\Sm}{{\mathbf S_{\bar t}}}
\newcommand{\hlp}{{\mathbf{\hat\ell}_+}}
\newcommand{\hlm}{{\mathbf{\hat\ell}_-}}

\newcommand{\hcvv}{{\hat c}_{VV}}
\newcommand{\hcaa}{{\hat c}_{AA}}
\newcommand{\hcva}{{\hat c}_{VA}}
\newcommand{\hcav}{{\hat c}_{AV}}

\def\xx1{\hat c_1}
\def\xy2{\hat c_2}
\def\xz3{\hat c_3}

\newcommand{\Hmu}{{\hat\mu}_t}
\newcommand{\Hd}{{\hat d}_t}
\newcommand{\stt}{\sigma_{t\bar t}}

\begin{document}

\begin{titlepage}
  \begin{flushright}
    TTK-24-11 
  \end{flushright}
  \vspace{0.01cm}
  
  \begin{center}
    {\LARGE \bf Binned top quark spin correlation and polarization observables for the LHC at 13.6~TeV. } \\
    \vspace{1.5cm}
    {\bf Werner Bernreuther}\,$^{a,}$\footnote{\tt
      breuther@physik.rwth-aachen.de}, 
    {\bf Long Chen}\,$^{b,}$\footnote{\tt longchen@sdu.edu.cn},
     {\bf and Zong-Guo Si}\,$^{b,}$\footnote{\tt zgsi@sdu.edu.cn}
    \par\vspace{1cm}
    $^a$Institut f\"ur Theoretische Teilchenphysik und Kosmologie, \\
    RWTH Aachen University,  52056 Aachen, Germany\\
    $^b$ School of Physics, Shandong University, Jinan, Shandong 250100, China
    \par\vspace{1cm}

   {\bf Abstract} \\
    \parbox[t]{\textwidth}
    {\small{ 
    We consider top-antitop quark $(\ttbar)$ production at the Large Hadron Collider (LHC) with
 subsequent decays into dileptonic final states. We use and investigate a  set of leptonic angular correlations and distributions
 with which  all the independent coefficient functions 
  of the top-spin dependent parts of the $\ttbar$ production spin density matrices can be experimentally probed.  
 We compute these  observables for the LHC center-of-mass energy 13.6~TeV within the Standard Model
 at next-to-leading order in the QCD coupling including the mixed QCD-weak corrections. 
 We determine also the $\ttbar$ charge asymmetry where we take in addition  also the mixed QCD-QED corrections into account. 
  In addition we analyze and compute possible new physics (NP) effects  on these observables within effective field theory
 in terms of a gauge-invariant  effective Lagrangian that contains the operators  up to mass dimension six that are relevant for
 hadronic $\ttbar$ production. First, we compute our observables inclusive in phase space. Then, in order to
 investigate which region in phase space has, for a specific observable,  a high NP sensitivity, we determine our 
  observables also in two-dimensional $(\mttbar,\cos\theta_t^*)$ bins, where $\mttbar$ denotes the $\ttbar$ invariant mass  and
$\theta_t^*$ is the top-quark scattering angle in the $\ttbar$ zero-momentum frame.
}}
 \end{center}

\vspace{1cm} \noindent
PACS number(s): 12.38.Bx, 13.88.+e, 14.65.Ha

\end{titlepage}

\setcounter{footnote}{0}
\renewcommand{\thefootnote}{\arabic{footnote}}
\setcounter{page}{1}

\section{Introduction}
\label{sec:intro}
The exploration of top-quark spin effects in hadronic top-quark pair production has become an
established tool for investigating Standard Model (SM) interactions and for searches of new physics (NP).
Evidence for  $\ttbar$ spin correlations was found first  by the D$\emptyset$ experiment \cite{D0:2011kcb} at the Tevatron  
 and they were first observed 
     by the ATLAS experiment  \cite{ATLAS:2012ao} at the Large Hadron Collider (LHC).
    Subsequently, the ATLAS and CMS experiments at the LHC   performed at center-of-mass-energies (c.m.)
    7 and 8 TeV, and more recently also at 13~TeV, a number of top spin correlation and polarization 
     measurements in the dileptonic and lepton plus jets final states, using various sets of observables
     \cite{ATLAS:2013gil,CMS:2013roq,ATLAS:2014aus,ATLAS:2014abv,CMS:2015cal,CMS:2016piu,ATLAS:2019zrq,CMS:2019nrx}.
     Recently, the ATLAS collaboration used top spin correlations to claim quantum entanglement in top quark pairs \cite{ATLAS:2023fsd}.
    This was confirmed by CMS \cite{cms:2024}.
    On the theory side, SM predictions for spin correlations and polarizations  were made at next-to-leading order (NLO) QCD including electroweak interactions
     for a number of $\ttbar$ spin correlation and top polarization observables in dileptonic and  lepton plus jets final states
     \cite{Bernreuther:2001rq,Bernreuther:2004jv,Bernreuther:2010ny,Frederix:2021zsh}. Radiative corrections to off-shell $\ttbar$ production
     and decay were determined in \cite{Bevilacqua:2010qb,Denner:2016jyo}. In \cite{Behring:2019iiv,Czakon:2020qbd} several $\ttbar$
     spin correlations were computed at next-to-next-to-leading order (NNLO) QCD for dileptonic final states.
     
A suggestion was made in \cite{Bernreuther:2015yna} to perform a comprehensive 
study of spin effects in hadronic $\ttbar$ production by measuring  all coefficients of the $\ttbar$ production spin density matrices. 
 For this aim the $gg, q{\bar q}\to\ttbar$ spin density matrices were decomposed in a suitable orthonormal basis and a set of spin observables
 was proposed that project onto the different coefficients of these density matrices. The $\ttbar$ spin correlation and $t$ and $\bar t$ polarization 
 observables were computed at NLO QCD including weak interaction corrections, and  possible NP effects on these 
 coefficients and observables were analyzed by using an effective NP Lagrangian.
 Other recent analyses of using $\ttbar$ spin correlations at the LHC for probing new physics include 
 \cite{Aguilar-Saavedra:2018ggp,Fabbrichesi:2022ovb,Severi:2022qjy,Maltoni:2024tul}.
 
 The suggestions of  \cite{Bernreuther:2015yna} were taken up first by the ATLAS experiment in \cite{ATLAS:2016bac} that measured at $\sqrt{s_{had}}=8$~TeV 
   a  subset  of the proposed spin observables and compared it with SM predictions \cite{Bernreuther:2015yna}.
  A more comprehensive analysis was performed by the CMS experiment \cite{CMS:2019nrx} at $\sqrt{s_{had}}=13$~TeV.
  Both experiments measured these spin observables for dileptonic final states inclusive in phase space and found agreement 
  with the SM results. Employing the NP computations of \cite{Bernreuther:2015yna} the CMS Collaboration used their data also to constrain
   anomalous NP top-quark couplings, in particular the anomalous chromomagnetic and chromoelectric dipole moments of the top quark \cite{CMS:2019nrx}. 
   Ref.~\cite{Severi:2022qjy} calculated for  a CP-even subset of the proposed spin observables in \cite{Bernreuther:2015yna} the NP contributions
   to quadratic order in the anomalous couplings and determined these observables differentially in phase space. 
 
  In this paper we extend the analysis of \cite{Bernreuther:2015yna} in several ways. We consider $\ttbar$ production and decay into dileptonic 
  final states at the present LHC c.m. energy $\sqrt{s_{had}}=13.6$~TeV. We compute the $\ttbar$ charge asymmetry $A_C$ and the set of
  top spin observables \cite{Bernreuther:2015yna} both in the SM at NLO QCD including weak interaction corrections and in the framework of an effective
  Lagrangian ${\cal L}_{\rm NP}$ describing new physics effects 
  in $\ttbar$ production \cite{Aguilar-Saavedra:2008nuh,Aguilar-Saavedra:2009ygx,Zhang:2010dr,Aguilar-Saavedra:2010uur,Degrande:2010kt}.
  We analyze two additional $\ttbar$ spin correlations besides those considered in \cite{Bernreuther:2015yna} that are useful 
  in pinning down two anomalous couplings of ${\cal L}_{\rm NP}$. First, we calculate our observables  inclusive in phase space. 
  Then, in order to explore which areas in phase space are particularly sensitive to the various anomalous couplings, 
  we choose a set of two-dimensional $(\mttbar,\cos\theta_t^*)$ bins, where $\mttbar$ denotes the $\ttbar$ invariant mass  and
$\theta_t^*$ is the top-quark scattering angle in the $\ttbar$ zero-momentum frame, and compute our observables bin by bin. 

Our paper is organized as follows. In section~\ref{sec:ortho} we briefly recapitulate the description of $\ttbar$ production and decay 
 in the spin-density matrix framework. Section~\ref{sec:effLag}
  contains the $SU(3)_c \times SU(2)_L \times U(1)_{Y}$ invariant effective NP Lagrangian  that we use.
  Our observables are introduced in section~\ref{sec:obs}. In section~\ref{sec:res13.6} we present our results for 
  the charge asymmetry $A_C$ and the spin observables, both inclusive in phase space and in two-dimensional 
   bins of $(\mttbar,\cos\theta_t^*)$. The results for the bins are given in detail in appendix~\ref{sec:AppA}.
  We conclude in section~\ref{sec:sumconc}.
  
\section{Formalism}
\label{sec:ortho}
We consider $\ttbar$ production  at NLO QCD including weak interaction corrections
 and subsequent semileptonic decays of $t$ and ${\bar t}$ quarks. At LO QCD top-pairs 
 are produced by $gg$ fusion and $\qqbar$ annihilation, and at NLO also $gq$ and $g{\bar q}$
 fusion contribute. We analyze $\ttbar$ production and decay in the so-called factorizable
  approximation and use the narrow width approximation $\Gamma_t/m_t \to 0$. The $t$
  and ${\bar t}$ spin degrees of freedom are fully taken into account. 
  In this approximation the squared matrix element $|{\cal M}_I|^2$ of the parton reactions
  $I \to F$ (where $F$ denotes here the dileptonic final state from $t$ and ${\bar t}$ decay,
  $F=b{\bar b} \ell^+ \ell'^- \nu_{\ell} {\bar\nu}_{\ell'}+X, (\ell= e,\mu,\tau)$) is of
  the form
  \begin{equation} \label{eq:genmatel}
   |{\cal M}_I|^2 \propto {\rm Tr}\left[\rho R^I {\bar \rho} \right] \, ,
  \end{equation}
 where $R^I$ denotes the density matrix that describes the production by one of the 
 above-mentioned initial parton reactions $I$ of on-shell $\ttbar$ pairs
in a specific spin configuration, and $\rho$ and ${\bar \rho}$ are the density matrices 
 that encode the semileptonic decays of polarized  $t$ and ${\bar t}$ quarks, see below. The trace extends over
  the $t$ and ${\bar t}$ spin indices.

We recall the structure of the $\ttbar$ production density matrices for the
    $2\to 2$ reactions
\begin{equation} \label{ggproc}
 g(p_1) + {g}(p_2) \to t(k_1, s_1) + {\bar t}(k_2, s_2) \, ,
 \end{equation}
\begin{equation} \label{qqproc}
 q(p_1) + {\bar q}(p_2) \to t(k_1, s_1) + {\bar t}(k_2, s_2) \, , 
\end{equation}
where $p_j, k_j$ and $s_1, s_2$ refer to the 4-momenta of the partons and to the
  spin 4-vectors of the $t$ and $\bar t$ quarks, respectively.
  The production density matrix $R^I$ 
 of  each of these reactions is defined by the squared modulus of the respective
  matrix element, averaged  over the spins and colors of $I$ and summed over the colors of $t, \bar t$.
 The structure of the $R^I$ $(I=gg, \qqbar)$ 
in the spin spaces of $t$  and $\bar{t}$ is as follows:
 \begin{equation} R^{I}=  f_I \left[ A^I 1\!{\rm l}\otimes 1\!{\rm l}+{\tilde B}^{I+}_{i}\sigma^i
\otimes 1\!{\rm l}+{\tilde B}^{I-}_{i}1\!{\rm l}
\otimes\sigma^i+{\tilde C}^I_{ij}\sigma^i\otimes\sigma^j \right] \, . \label{Rot} 
\end{equation}
The first (second) factor in the tensor products of the $2\times 2$
unit matrix $1\!{\rm l}$ and of the Pauli matrices $\sigma^i$ refers to the $t$
 $(\bar{t})$ spin space.
 The prefactors are given to LO QCD by
 \begin{equation}
  f_{gg} =  \displaystyle{\frac{(4\pi\alpha_s)^2}{N_c (N_c^2-1)} \, , \qquad f_{q\bar q}=\frac{(N_c^2-1)(4\pi\alpha_s)^2}{N_c^2} } \, , \nonumber 
\end{equation}
 where $N_c=3$ denotes the number of colors. 

The functions ${\tilde B}^{I\pm}_{i}$ and ${\tilde C}^I_{ij}$ can
be further decomposed, using an orthonormal basis which we choose as in \cite{Bernreuther:2015yna}.  
 The top-quark direction of flight in the $\ttbar$ zero-momentum frame (ZMF) is denoted by $\hk$,
 and $\hp={\bf\hat p}_1$ denotes the direction of one of the incoming partons in this frame. 
  A right-handed orthonormal
   basis is obtained as follows:
\begin{align} \label{orhtoset}
	\bigl\{\hr, \, \hk, \, \hn \bigr\}: \qquad \hr = \frac{1}{r}\bigl(\hp - y\hk\bigr),\quad
	\hn = \frac{1}{r}\bigl(\hp \times \hk \bigr),&\qquad
	y = \hk \cdot \hp, \; r = \sqrt{1-y^2}. 
\end{align}
 Using rotational invariance we decompose
  the 3-vectors ${\bf\tilde B}^{I\pm}$ and the
 $3\times 3$ matrices ${\tilde C}^I_{ij}$ (which have a symmetric and antisymmetric part with 6 and
 3 entries, respectively) with respect to the  basis \eqref{orhtoset}:
\begin{align}
	{\tilde B}^{I \pm}_i &=    b^{I\pm}_r\, \hat r_i \;+\; b^{I\pm}_k\, \hat k_i  \;+\; b^{I\pm}_n \, \hat n_i \, , \label{eq:newB} \\
	{\tilde C}^I_{ij} &=\;  c^I_{rr}\,\hat r_i \hat r_j \;+\; c^I_{kk}\,\hat k_i \hat k_j \;+\; c^I_{nn}\,\hat n_i \hat n_j \notag \\
                    & +\; c^I_{rk}\,\bigl(\hat r_i \hat k_j \;+\; \hat k_i \hat r_j \bigr) 
                +\; c^I_{kn}\,\bigl(\hat k_i \hat n_j \;+\; \hat n_i \hat k_j \bigr) 
              \;  +\; c^I_{rn}\,\bigl(\hat r_i \hat n_j \;+\; \hat n_i \hat r_j \bigr) \notag\\
	&+\; \epsilon_{ijl}\,\bigl( c^{I}_r \, \hat r_l \;+\; c^{I}_k \, \hat k_l \;+\; c^{I}_n  \hat n_l   \bigr) \, . \label{eq:newC}
\end{align}
 The coefficients  $b^{I\pm}_v$,  $c^I_{vv^{\prime}}$  are functions of  the partonic c.m. energy  squared, $\hat s$,
  and of $y=\hk \cdot \hp$   which is equal to the  cosine of the top-quark scattering angle in the c.m. frame of the initial partons.
 Notice that the terms in the antisymmetric part of \eqref{eq:newC} can be written as follows:
  \begin{equation} \label{antiCnew}
  c^{I}_r\epsilon_{ijl}\hat r_l =c^{I}_r(\hat k_i \hat n_j -\hat n_i \hat k_j), \quad 
  c^{I}_k\epsilon_{ijl}\hat k_l =c^{I}_k(\hat n_i \hat r_j -\hat r_i \hat n_j), \quad  
  c^{I}_n\epsilon_{ijl}\hat n_l =c^{I}_n(\hat r_i \hat k_j -\hat k_i \hat r_j).  
  \end{equation}
  Bose symmetry of the initial $gg$ state implies that the matrix $R^{gg}$ must satisfy
 \begin{equation}
  R^{gg}(-{\bf p}, {\bf k}) = R^{gg}({\bf p}, {\bf k}) \, .
  \label{bose}
  \end{equation}
  If CP invariance holds then
  \begin{equation}
   R^{I}_{\alpha_1 \alpha_2,\beta_1 \beta_2}({\bf p}, {\bf k}) = R^{I}_{\beta_1 \beta_2, \alpha_1 \alpha_2}({\bf p}, {\bf k}) \, ,
   \qquad I= g g, \, q {\bar q} \, .
   \label{CPinv}
  \end{equation}
The conditions \eqref{bose} and \eqref{CPinv} imply transformation properties of the  coefficient functions 
$b^{I\pm}_v$,  $c^I_{vv^{\prime}}$  defined in \eqref{eq:newB} and \eqref{eq:newC}. These properties are listed,
 together with the implications of parity invariance, in detail in Table~1 of \cite{Bernreuther:2015yna}. 
 
We briefly recall the structure of the decay density matrices $\rho$, ${\bar\rho}$ that we use. 
In this paper we concentrate on semileptonic top-quark decays. At NLO QCD we
have 
\begin{equation} \label{semilt}
 t \to b \ell^+\nu_{\ell}, \; b \ell^+\nu_{\ell} g \, ,  
\end{equation}
where $\ell=e,\mu,\tau$. Considering a fully polarized ensemble of top quarks in the top rest frame and integrating over all energy and angular variables in the
decay matrix element, except over the angle $\theta$ between the polarization vector of the top quark and the direction
of flight of the charged lepton $\ell^+$, one obtains a decay distribution of the form 
$d\Gamma_{\ell}/d\cos\theta=\Gamma_{\ell}\left(1+\kappa_{\ell}\cos\theta\right)/2$, where $\Gamma_{\ell}$ is the partial width of the respective
semileptonic decay. From this decay distribution one obtains the respective normalized one-particle inclusive $t$-decay density matrix
\begin{equation} \label{eq:tdecdens}
  \rho = \frac{1}{2}\left( 1\! + \kappa_{\ell}{\mathbf\sigma} \cdot \hlp \right) \, .
\end{equation}
The factor $\kappa_{\ell}$ is the top-spin analyzing power of the charged lepton.
 Its value is $\kappa_\ell=0.999$ at NLO QCD \cite{Czarnecki:1990pe,Brandenburg:2002xr}.
For semileptonic ${\bar t}$ decays the respective normalized decay density matrix is
  \begin{equation} \label{eq:tbardecdens}
  {\bar\rho} = \frac{1}{2}\left( 1\! - \kappa_{\ell}{\mathbf\sigma} \cdot \hlm \right) \, .
\end{equation}
Eqs.~\eqref{eq:tdecdens} and~\eqref{eq:tbardecdens} will be used in  \eqref{eq:genmatel}
 for the computation of the spin observables
of section~\ref{sec:obs}. The proportionality factor in  \eqref{eq:genmatel} contains, in
the narrow width approximation, the branching fractions of semileptonic $t$ and ${\bar t}$ decay.

\section{Effective NP Lagrangian.}
\label{sec:effLag}
Assuming that new physics (NP) effects in hadronic $\ttbar$ production and decay are 
characterized by a mass scale $\Lambda$ that is significantly larger than the moduli of the kinematic invariants of
the $\ttbar$ production and decay processes, one may describe these (non-resonant) effects by a local effective Lagrangian ${\mathcal L}_{\rm NP}$
that involves the SM degrees of freedom and respects the SM symmetries. Respective analyses include
\cite{Aguilar-Saavedra:2008nuh,Aguilar-Saavedra:2009ygx,Zhang:2010dr,Aguilar-Saavedra:2010uur,Degrande:2010kt}.

The gauge-invariant operators with ${\rm dim}{\cal O}\le 6$ relevant for $\ttbar$ production 
that involve gluon fields are  \cite{Aguilar-Saavedra:2008nuh,Zhang:2010dr,Degrande:2010kt}
  \begin{align}   
  \mathcal O_{gt} &= \bigl[ \bar t_R \gamma^{\mu} T^a D^{\nu} t_R \bigr] \,G^a_{\mu\nu} \, , \label{eq:Ogt}\\
	      \mathcal O_{gQ} &= \bigl[ \bar Q_L \gamma^{\mu} T^a D^{\nu} Q_L \bigr] \,G^a_{\mu\nu} \label{eq:OgQ} \, ,\\
	      \mathcal O_{\text{CDM}} &= \bigl[ \bigl({\tilde\Phi}\bar Q_L \bigr) \sigma^{\mu\nu} T^a t_R \bigr] \,G^a_{\mu\nu} \, . \label{eq:Hg}
   \end{align}   
    Here $Q_L=(t_L, b_L)$ is the left-handed third generation doublet and ${\tilde\Phi}=i\sigma_2\Phi^\dagger=(\phi_0^*,-\phi_-)$ is the charge-conjugate
     Higgs doublet field. Moreover, $D_{\mu} = \partial_{\mu} + i g_s T^a G^a_{\mu}$ and 
	 $G^a_{\mu\nu} = \partial_{\mu} G^a_{\nu} - \partial_{\nu} G^a_{\mu} - g_s f^{abc} G^b_{\mu} G^c_{\nu}$ is the gluon field strength tensor.
	  Furthermore,  $T^a$ are the generators of $\text{SU}(3)_c$ in the fundamental representation, with ${\rm tr}(T^aT^b)=\delta_{ab}/2$. 

	  The sums $\mathcal O_{gt}+ \mathcal O_{gt}^\dagger$
	   and $\mathcal O_{gQ}+\mathcal O_{gQ}^\dagger$
	   are given by linear combinations of four-quark operators  as can be shown using the equation of motion for the gluons \cite{Degrande:2010kt}.
	   These linear combinations of  four-quark operators are redundant in our case, because they are included in the 
	   set of four-quark operators given below that we use.
	  The combinations $\mathcal O_{gt}-\mathcal O_{gt}^\dagger$ and $\mathcal O_{gQ}-\mathcal O_{gQ}^\dagger$
	  cannot be expressed in terms of the  four-quark operators below. With these two combinations and \eqref{eq:Hg} one can construct an 
	  hermitean effective Lagrangian which reads, after spontaneous symmetry breaking,
	       $\langle \Phi\rangle=v/\sqrt{2}$, ($v\simeq 246$ GeV), restriction to the top-quark gluon sector, and using operators with definite P and CP properties:
	       \begin{equation} \label{eq:Lefftg}
	       {\mathcal L}_{\rm NP, g}= -\frac{g_s}{2 m_t}\left[\Hmu {\bar t}\sigma^{\mu\nu}T^a t G^a_{\mu\nu} +
	        \Hd {\bar t}i\sigma^{\mu\nu}\gamma_5 T^a t G^a_{\mu\nu} \right] \, .
	        \end{equation}	       
In the convention used in \eqref{eq:Lefftg} we used the top-quark mass $m_t$ for setting the mass scale. The real and dimensionless coupling parameters $\Hmu$ and $\Hd$
 are, respectively, the chromomagnetic and chromoelectric dipole moment of the top quark.
 
In  \cite{Bernreuther:2015yna} we used  an effective gluon Lagrangian that is $SU(3)_c \times U(1)_{em}$
  invariant, but not $SU(3)_c \times SU(2)_L \times U(1)_{Y}$ invariant, for reasons of a more agnostic, phenomenological analysis. 
  Besides  \eqref{eq:Lefftg} it contains two additional CP-odd operators that contribute to  P- and CP-odd spin correlations 
  and to a P-even, CP-odd polarization observable, respectively. These observables were measured in \cite{CMS:2019nrx}   
  and the respective coupling parameters ${\hat c}(- -)$ and  ${\hat c}(- +)$
  were constrained. In this paper we will not use these operators, but stick to the $SU(3)_c \times SU(2)_L \times U(1)_{Y}$ invariant framework. 
	       
There are a number of gauge-invariant ${\rm dim}{\cal O}= 6$ four-quark operators that generate non-zero tree-level interference terms 
with the $q{\bar q}\to \ttbar$  QCD amplitude. Assuming universality of the new interactions with respect to the light quarks $q\ne t$ and 
considering only operators with $u,d$ quarks in the initial state
 that interfere with the tree-level $\qqbar \to \ttbar$ QCD matrix elements, seven gauge-invariant operators  remain \cite{Zhang:2010dr,Degrande:2010kt}.
It is useful  to combine these seven operators such that 
	   one obtains four isospin-zero operators with definite P and C properties  and 
	   three isospin-one operators \cite{Degrande:2010kt}. 
	   The resulting NP effective Lagrangian involving the $u,d$ quarks reads (as above we use $m_t$ for setting the mass scale):
	    	   	     \begin{eqnarray} \label{eq:Leffqua}
	       {\mathcal L}_{\rm NP, q}= {\mathcal L}_{\rm NP, 0} + {\mathcal L}_{\rm NP, 1}  \, ,
	       \end{eqnarray}
	       where the isospin-zero part is
             \begin{eqnarray} \label{eq:Leffqu0}
	       {\mathcal L}_{\rm NP, 0}=\frac{g_s^2}{2 m_t^2} \sum\limits_{I,J=V,A} {\hat c}_{IJ}  \mathcal O_{IJ} \, ,
	       \end{eqnarray}
	       and
	       \begin{eqnarray} \label{eq:Opqu0}
	\mathcal O_{VV} = ({\bar q}\gamma^\mu T^a q)({\bar t}\gamma_\mu T^a t) \, , & \qquad \mathcal O_{AA} = ({\bar q}\gamma^\mu T^a \gamma_5 q)({\bar t}\gamma_\mu \gamma_5 T^a t) \, ,	\\
	\mathcal O_{VA} = ({\bar q}\gamma^\mu T^a  q)({\bar t}\gamma_\mu \gamma_5 T^a t) \, , & \qquad \mathcal O_{AV} = ({\bar q}\gamma^\mu T^a \gamma_5 q)({\bar t}\gamma_\mu T^a t) \, .
	  \end{eqnarray}  
	   Here and in the following $q=(u,d)$ denotes the isospin doublet.
	  	   The isospin-one contribution can be represented in the form
	   \begin{eqnarray} \label{eq:Leffqu1}
	       {\mathcal L}_{\rm NP, 1}=\frac{g_s^2}{4 m_t^2} \sum\limits_{i=1}^3 {\hat c}_{i} \mathcal O^1_i \, ,
	       \end{eqnarray}
	       where
	        \begin{eqnarray} \label{eq:Opqu1}
	 \mathcal O^1_1 = & ({\bar q}\gamma^\mu T^a \sigma_3 q)({\bar t}\gamma_\mu T^a t) + ({\bar q}\gamma^\mu \gamma_5 T^a  \sigma_3 q)({\bar t}\gamma_\mu T^a t) \, ,\\
	\mathcal O^1_2 =  & ({\bar q}\gamma^\mu \gamma_5 T^a \sigma_3 q)({\bar t}\gamma_\mu \gamma_5 T^a t) - ({\bar q}\gamma^\mu \gamma_5 T^a  \sigma_3 q)({\bar t}\gamma_\mu T^a t) \, ,\\
	 \mathcal O^1_3 =  & ({\bar q}\gamma^\mu T^a \sigma_3 q)({\bar t}\gamma_\mu \gamma_5 T^a t) + ({\bar q}\gamma^\mu \gamma_5 T^a  \sigma_3 q)({\bar t}\gamma_\mu T^a t) \, .
	 \end{eqnarray}
  In the isospin-one case it is not possible to combine the operators such that they have definite properties
	     with respect to C and P.
	     
In summary we use the effective NP Lagrangian	     
 \begin{equation}\label{eq:LNPsum}
              {\mathcal L}_{\rm NP} = {\mathcal L}_{\rm NP, g} + {\mathcal L}_{\rm NP, q}
       \end{equation}
       that contains the real, dimensionless coupling parameters $\Hmu$, $\Hd$,  ${\hat c}_{IJ}$ $(I,J=V,A)$, and 
       ${\hat c}_1, {\hat c}_2, {\hat c}_3.$ 
       The NP contributions to the coefficients of the $\ttbar$ spin density matrices \eqref{Rot} induced by interference
        with the tree-level QCD amplitudes of $gg, q{\bar q}\to \ttbar$ are listed in the Appendix of \cite{Bernreuther:2015yna}.
        In the following we will stick to these dependencies to first order in the anomalous couplings. This is justified a posteriori
        by the results \cite{CMS:2019nrx} of the CMS experiment. The experimental 
        constraints on the dimensionless anomalous couplings of \eqref{eq:LNPsum} signify that they are markedly smaller than one.
       	     
	A remark on NP contributions to top-quark decay $t\to b\ell\nu_{\ell}$  is in order. 
	The top-quark decay vertex $t\to W b$ may be affected by new physics interactions,
	but the upper bounds on the respective anomalous couplings inferred from measurements of 
	the $W-$boson helicity fractions \cite{Fabbrichesi:2014wva,Cao:2015doa}
	show that these effects
	 are very small if non-zero. 
	In this paper, we consider $\ttbar$ production and decay
	 in the dileptonic channel, $pp \to \ttbar X\to \ell \ell' X$. We use as top-spin analyzers the charged lepton from $W$
	  decay and we analyse only charged-lepton angular observables that are inclusive in the lepton energies.
	  These observables are not affected by anomalous couplings
 from top-quark decay if a linear approximation is justified, that is, if these couplings are 
 small \cite{Rindani:2000jg,Grzadkowski:1999iq,Grzadkowski:2002gt,Godbole:2006tq}. As just mentioned this is the case.
 Thus for the observables that we analyse in this paper only contributions to $\ttbar$ production matter as far as NP effects are concerned.

\section{Observables}
\label{sec:obs}

We consider $\ttbar$ production at the LHC for the present center-of-mass energy
$\sqrt{s}_{\rm had}=13.6$~TeV. We compute the $\ttbar$ cross section and the $\ttbar$ charge
asymmetry $A_C$ defined in \eqref{eq:LHCchaas} below in the SM and determine, in addition, the contributions from ${\mathcal L}_{\rm NP}$.
Then we focus on the dileptonic $\ttbar$ decay channels and investigate a number of spin correlation and polarization observables.
First we perform our computations within the SM and to first order 
 in  ${\mathcal L}_{\rm NP}$ inclusive in phase space. 
 Because future experimental investigations at the LHC aim at 
more detailed analyses, we then determine these observables in two-dimensional bins of the $\ttbar$ invariant mass
 and  the cosine of the top-quark scattering angle in the $\ttbar$ zero-momentum frame (ZMF) that will be specified below.
 We do not apply acceptance cuts because  experiments usually unfold their data for comparison with 
 theoretical (top-spin) predictions \cite{ATLAS:2016bac,CMS:2019nrx}.

 We use the LHC $\ttbar$ charge asymmetry defined by
       \begin{equation} \label{eq:LHCchaas}
     A_C = \frac{ \sigma(\Delta|y|>0) - \sigma(\Delta|y|<0)}{ \sigma(\Delta|y|>0) + \sigma(\Delta|y|<0)}  
    \end{equation}
     where $\Delta|y|=|y_t|-|y_{\bar t}|$ is the difference of the moduli of the $t$ and $\bar t$ rapidities
 in the laboratory frame.

 For the dileptonic final states 
 \begin{equation}\label{ppdilep}
p p \to \ttbar + X\to \ell^ + \ell'^-  + \ \text{jets}  + X\, ,
\end{equation}
 we consider the well-known polar angle double distributions \cite{Bernreuther:2001rq,Bernreuther:2004jv} for a choice of reference axes $\ha, \, \hb$:
\begin{align}
	\frac{1}{\sigma} \frac{d\sigma}{d\cos\theta_+ d\cos\theta_-} = \frac{1}{4}\Bigl( 1 + B_1 \, \cos\theta_+ +  B_2 \, \cos\theta_- 
	- C\, \cos\theta_+\cos\theta_- \, \Bigr) \, ,
	\label{eq:doublediff}
\end{align}
where
\begin{align}
	\cos\theta_{+} = \hlp \cdot \ha \, , \qquad 	\cos\theta_{-} =\hlm \cdot \hb \, ,
\end{align}
and, as above, the unit vectors $\hlp$, $\hlm$ are the $\ell^+$ and $\ell'^-$ directions of flight in the $t$ and $\bar t$ rest frames, respectively.
The coefficients $B_1$, $B_2$ and $C$ signify the $t$, ${\bar t}$ polarizations and $\ttbar$ spin correlations, respectively.
As we apply no acceptance cuts
 on the final states and consider only factorizable radiative corrections the coefficients in \eqref{eq:doublediff} can be 
related to the expectation values of the spin observables at the level of the intermediate top quarks.
 We have \cite{Bernreuther:2004jv}
\begin{equation}
 C(a,b) = \kappa_\ell^2 \langle 4 (\Sp\cdot\ha) (\Sm\cdot\hb)\rangle =
\kappa_\ell^2  \langle  {\mathbf\sigma}\cdot\ha \otimes {\mathbf\sigma}\cdot\hb\rangle \, . 
\label{eq:Cab}
\end{equation}
 Here $\Sp$ and $\Sm$ denote the $t$ and ${\bar t}$ spin operators and ${\sigma_i}$ are the Pauli matrices. The 
value of  $\kappa_\ell$ is listed below Eq.~\eqref{eq:tdecdens}.

 The coefficients $B_{1,2}$ in \eqref{eq:doublediff} are given by 
\begin{equation} \label{c4:relpol}
  B_1(\ha) = P(\ha)~\kappa_\ell \, , \qquad B_2(\hb) = - {\overline P}(\hb)~\kappa_\ell \, ,
\end{equation}
where $P$, ${\overline P}$ denote the polarization degrees of the  $t$ and $\bar t$ ensembles in $\ttbar$ events 
  with respect  to the reference axes  $\ha, \, \hb$:  
\begin{equation} \label{eq:polde}
P(\ha)=\langle 2 \Sp\cdot\ha\rangle \, , 
\qquad {\overline P}(\hb)=\langle 2 \Sm\cdot\hb\rangle \, .
\end{equation}
 The relative signs  in front of the coefficients $B_{1,2}$ in the distribution \eqref{eq:doublediff}
 are chosen such that in a CP-invariant theory and for 
  the choice $\ha  = - \hb$:
\begin{align}
	B_1 = B_2 \, . \label{B1B2}
\end{align}

For the choice of reference axes $\ha$ and $\hb$ at the hadron level one cannot use 
the orthonormal basis at the parton level 
 introduced in section \ref{sec:ortho}, because the incoming quark will be either in the right-moving
 or in the left-moving proton. As in \cite{Bernreuther:2015yna} we choose the following set:
 We use the unit vector $\hk$ which is the top quark direction
 of flight in the $\ttbar$ ZMF.
 Moreover, we use the direction of  one of the proton beams in the
  laboratory frame, ${\hp}_p$, and define unit vectors  ${\hr}_p$ and ${\hn}_p$ as follows: 
 \begin{align}
  {\hp}_p =(0,0,1) \, , & \quad {\hr}_p =\frac{1}{r_p}({\hp}_p -y_p \hk) \, , & \quad {\hn}_p =\frac{1}{r_p} ({\hp}_p \times \hk) \, , 
   \label{proONB} \\
    & \quad y_p = {\hp}_p\cdot \hk =\cos\theta_t^*\, , & \quad r_p=\sqrt{1-y^2_p}  \, . \nonumber
   \end{align}
The angle $\theta_t^*$ is the top-quark scattering angle 
in the $\ttbar$ ZMF. Only in the case of $2\to 2$ parton reactions and if the incoming parton 1 is parallel to  ${\hp}_p$, the unit vectors  defined in \eqref{orhtoset}
 are the same as those in \eqref{proONB}. The set \eqref{proONB} defines our choice  
  of reference axes $\ha$ and $\hb$ which we  list in Table~\ref{tab:axes}. 
 The factors $\text{sign}(y_p)$ are required because
of the Bose symmetry of the initial $gg$ state. 
\begin{table}[!t]
\begin{center}
	\caption{Choice of reference axes at the hadron level. The unit vectors ${\hn}_p$, ${\hr}_p$ and the 
	variable $y_p$ are defined in \eqref{proONB}.}
	  \vspace{1mm}
	  \label{tab:axes}
  {\renewcommand{\arraystretch}{1.2}
\renewcommand{\tabcolsep}{0.2cm}
    \begin{tabular}{l c c c } \hline
	& label    	&	~~~$\ha $ ~~~		&	~~~$\hb$~~~ \\ \hline\hline
transverse &	n		&	~~~$\text{sign}(y_p)\; {\hn}_p$~~~	&	~~~$-\text{sign}(y_p)\; {\hn}_p$~~~ \\
r axis & 	r		&	$\text{sign}(y_p)\; {\hr}_p$	&	$-\text{sign}(y_p)\; {\hr}_p$ \\
helicity &	k		&	$\hk$		&	$-\hk$ \\ \hline

	\end{tabular} }
\end{center}	
\end{table}
In the following the label  $(a,\, b)$ refers to the choice of reference axes
 $\ha$ and $\hb$ from  Table~\ref{tab:axes}. The correlation coefficient  $C$   associated  
 with this choice of axes is denoted by $C(a, \, b)$  as in \eqref{eq:Cab} and is called $\ttbar$ spin correlation for short.
 The analogous labeling applies to  $B_1(a)$    and   $B_2(b)$ which we refer to as $t$ and ${\bar t}$ polarization with respect
 to the chosen axis.
 
   \begin{table}[!htb]
   \begin{center}
\caption{The spin correlations and polarizations of (\ref{eq:doublediff}),  
         and sums and differences  for different choices of reference axes. The unit vectors associated with 
         the labels $k^*$ and $r^*$  are defined in \eqref{eq:kstardef} and \eqref{eq:rstardef}.  }
	\label{tab:C-B-coeff}
	  \vspace{1mm}
  {\renewcommand{\arraystretch}{1.2}
\renewcommand{\tabcolsep}{0.2cm}
	 \begin{tabular}{l l l } \hline
		Correlation		&	                	&	sensitive to \\\hline\hline
		$C(n,\,n)$ 		&	$c^I_{nn}$		&	P-even, CP-even\\
		$C(r,\,r)$ 		&	$c^I_{rr}$		&	P-even, CP-even\\
		$C(k,\,k)$ 		&	$c^I_{kk}$		&	P-even, CP-even\\
		$C(r,\,k)+C(k,\,r)$ 	&	$c^I_{rk}$		&	P-even, CP-even\\
		$C(n,\,r)+C(r,\,n)$ 	&	$c^I_{rn}$                &       P-odd, CP-even, absorptive\\
		$C(n,\,k)+C(k,\,n)$ 	&	$c^I_{kn}$		&	 P-odd, CP-even,  absorptive\\
		$C(r,\,k)-C(k,\,r)$ 	&	$c^{I}_n$	&	 P-even, CP-odd, absorptive \\
		$C(n,\,r)-C(r,\,n)$ 	&	$c^{I}_k$	&	 P-odd, CP-odd\\
		$C(n,\,k)-C(k,\,n)$ 	&	$c^{I}_r$	&	 P-odd, CP-odd\\
		$C(k,\,k^*)$            &       $c^{I}_{kk}$    &      P-even, CP-even \\
		$C(r^*,\,k)+C(k,\,r^*)$ &       $c^I_{rk}$      &      P-even, CP-even \\
		$B_1(n) + B_2(n)$ 	&	$b^{I+}_{n} + b^{I-}_{n}$	&	P-even, CP-even, absorptive\\
		$B_1(n) - B_2(n)$ 	&	$b^{I+}_{n} - b^{I-}_{n}$ 	&	 P-even, CP-odd\\
		$B_1(r) + B_2(r)$ 	&	$b^{I+}_{r} + b^{I-}_{r}$ 	&	 P-odd, CP-even \\
		$B_1(r) - B_2(r)$ 	&	$b^{I+}_{r} - b^{I-}_{r}$ 	&	 P-odd, CP-odd, absorptive\\
		$B_1(k) + B_2(k)$ 	&	$b^{I+}_{k} + b^{I-}_{k}$ 	&	 P-odd,CP-even \\
		$B_1(k) - B_2(k)$ 	&	$b^{I+}_{k} - b^{I-}_{k}$ 	&	 P-odd, CP-odd, absorptive\\
	        $B_1(r^*) + B_2(r^*)$ 	&	$b^{I+}_{r} + b^{I-}_{r}$ 	&	 P-odd, CP-even \\
		$B_1(r^*) - B_2(r^*)$ 	&	$b^{I+}_{r} - b^{I-}_{r}$ 	&	 P-odd, CP-odd, absorptive	\\
	        $B_1(k^*) + B_2(k^*)$ 	&	$b^{I+}_{k} + b^{I-}_{k}$ 	&	 P-odd,CP-even \\
		$B_1(k^*) - B_2(k^*)$ 	&	$b^{I+}_{k} - b^{I-}_{k}$ 	&	 P-odd, CP-odd, absorptive\\ \hline	
		\end{tabular} }
 \end{center}
\end{table}

 Table~\ref{tab:C-B-coeff} contains the set of spin correlations and polarizations that we consider. The second column shows to which coefficient
 of the $gg$ and $\qqbar$ spin density matrices the respective observable is sensitive. The third column indicates the P- and CP-symmetry
 properties of the observables.\footnote{We recall that, strictly speaking, a classification at the hadron level 
 with respect to $CP$ is not possible, because the initial $p p$ state is not a CP eigenstate. The third column of table~\ref{tab:C-B-coeff} refers to
  the P- and CP-symmetry properties of the coefficients of the $gg$ and $\qqbar$ spin density matrices \cite{Bernreuther:2015yna}.} The label ``absorptive''
  means that the respective observable is generated by absorptive parts in the scattering matrix.
 
 Apart from computing the observables of Table~\ref{tab:C-B-coeff} in the SM we determine also their sensitivity to the couplings of the 
 effective NP Lagrangian \eqref{eq:LNPsum}.
 It contains the following
 dimensionless, real parameters: ${\hat\mu}_t$, ${\hat d}_t$, ${\hat c}_{VV}$,  ${\hat c}_{AA}$,
 ${\hat c}_{AV}$,  ${\hat c}_{VA}$, ${\hat c}_1$, ${\hat c}_2$, ${\hat c}_3$.
 The appendices A.1 and A.2 of  \cite{Bernreuther:2015yna} show on which parameters a specific coefficient of the $gg$ and 
 $\qqbar$ spin density matrices depends.\footnote{There are typos in the following two coefficients in Appendix  A.2 of 
 \cite{Bernreuther:2015yna}.
 The overall sign of the first term of $c_{rr}$ must be positive, instead of negative. The second term in
 the coefficient $c_k$ should read $\beta(-1+y^2)\Hd/4$. The numerical results of \cite{Bernreuther:2015yna} are not affected by these typos. }
 This dependence  determines the
 dependence on specific NP parameters of $\sigma_{\tt}$, $A_C$ and the spin observables $B$ and $C$.
 It turns out that in order to significantly increase the sensitivity to some of these parameters,
 it is useful to introduce, in addition to those of Table~\ref{tab:axes}, another set of reference axes \cite{Bernreuther:2015yna}
 to which we assign the labels $k^*$ and $r^*$:
\begin{align}
 k^*: \quad & {\ha} = {\rm sign}(\Delta|y|)~{\hk} \, , \quad {\hb} = -{\rm sign}(\Delta|y|)~{\hk}\, , \label{eq:kstardef}\\
 r^*: \quad & {\ha} = {\rm sign}(\Delta|y|)~\text{sign}(y_p)~{\hr}_p \, ,\quad   
 {\hb} = -{\rm sign}(\Delta|y|)~\text{sign}(y_p)~{\hr}_p \, . \label{eq:rstardef}
\end{align}
Here  $\Delta|y|$  denotes the difference of the moduli of the $t$ and $\bar t$ rapidities
 in the laboratory frame as defined below \eqref{eq:LHCchaas}.  With these vectors, one may consider the spin observables
 \begin{equation} \label{eq:sumBkrstar}
 C(k,k^*), \quad C(r^*,k), \; C(k,r^*),  \qquad B_1(k^*), \;  B_2(k^*) \, , \qquad B_1(r^*), \;  B_2(r^*)   \, ,
\end{equation} 
 respectively their sums and differences, see Table~\ref{tab:C-B-coeff}. The sum of $B_{1,2}(k^*)$ and of  $B_{1,2}(r^*)$
 is sensitive to the  NP contributions
   from the operator   $\mathcal O_{AV}$ and from a P-odd combination of the operators  $\mathcal O^1_i$,
   while the sum of
    $B_{1,2}(k)$ and $B_{1,2}(r)$ projects onto the contributions of $\mathcal O_{VA}$ and  $\mathcal O^1_3$. 
    The spin correlations $C(k,k^*)$ and  $C(r^*,k)+ C(k,r^*)$, which were not computed in \cite{Bernreuther:2015yna}, project onto different
    regions of phase space than $C(k,k)$ and  $C(r,k) + C(k,r)$. While the latter are sensitive to $\hcvv$, $\xx1$, and $\Hmu$,
    the former probe the couplings $\hcaa$ and $\xy2$. The P- and CP-odd correlations $C(n,r)-C(r,n)$ and $C(n,k)-C(k,n)$ 
    are equivalent to CP-odd triple correlations, cf. \cite{Bernreuther:2015yna}, and they probe the chromoelectric dipole moment
     of the top quark.
    
    Table~\ref{tab:C-B-coeff} contains a number of P-odd observables that require absorptive parts. In the SM they result from absorptive parts 
    of weak-interaction contributions and are very small and we do not compute them. There are also no such contributions from
     our ${\cal L}_{\rm NP}$ at tree-level. The observables $B_1(n)-B_2(n)$ and  $C(r,k)-C(k,r)$ are P-even, but CP-odd. 
      The latter requires in addition absorptive parts. Neither SM nor NP interactions from \eqref{eq:LNPsum} contribute to these observables.

 We close this section with a remark on the opening angle distribution \cite{Bernreuther:2001rq,Bernreuther:2004jv} $\sigma^{-1}d\sigma/d\cos\varphi=(1-D\cos\varphi)/2$
  where $\cos\varphi =\hlp \cdot \hlm$ is the scalar product of the two lepton directions determined in their parent $t$ and ${\bar t}$ rest
  frames. Measurements by ATLAS and CMS have shown (see, for instance, \cite{ATLAS:2016bac,CMS:2019nrx}) that
  this distribution is highly sensitive to $\ttbar$ spin correlations.
  It can be obtained from the diagonal spin correlation coefficients. Using that the vectors defined in Table~\ref{tab:axes}
  form orthonormal sets one gets \cite{Bernreuther:2015yna}
  \begin{equation}
  \label{eq:Dcor}
     D = -\frac{1}{3} \left[ C(r,r) + C(k,k) + C(n,n)\right] \, .
  \end{equation}
 The opening angle distribution can be determined with this formula from the diagonal correlations
 that will be computed in the next section.
 
\section{Results for 13.6 TeV}
\label{sec:res13.6}
We compute the cross section, the charge asymmetry $A_C$,
and the expectation values of the above spin observables for $ p p$ collisions at the c.m. energy 
of 13.6~TeV.
 As already emphasized above we do not apply acceptance cuts on the final states, because experiments compare
 with theory predictions by correcting their measurements to the parton level and extrapolating to the full phase space
 (see, e.g., \cite{ATLAS:2016bac,CMS:2019nrx}).
 We use the CT18 NLO parton distribution functions\cite{Hou:2019efy}. 
 This set provides also the NLO QCD coupling $\alpha_s$ in the ${\overline{\rm MS}}$ scheme.
 We use the on-shell top-quark mass $m_t = 172.5 ~\rm GeV.$
 Moreover, we use $\Gamma_t  =1.3  \ {\rm GeV}$, $m_Z=91.2$ GeV,   $m_W=  80.4 \ {\rm GeV}, 
 \Gamma_W = 2.09 \ {\rm GeV}$, $m_H=125$ GeV, and $\alpha(m_t)=0.008$.
 We perform our computations for three values of the renormalization
  and factorization scale $\mu_R=\mu_F = \mu$, namely for $\mu= m_t/2, m_t, 2 m_t$. 
 
 First we compute our observables inclusively, i.e., by integrating over the complete phase space. 
 Then we determine their values in two-dimensional bins of the $\ttbar$ invariant mass $\mttbar$
  and $y_p =\cos\theta_t^*$, where $\theta_t^*$ is the top-quark scattering angle in the $\ttbar$ 
  ZMF. 
 
 We choose the four $\mttbar$ intervals 
 \begin{eqnarray} \label{mttbin}
  0\leq \mttbar \leq 450~{\rm GeV}, \quad 450~{\rm GeV} < \mttbar \leq 600~{\rm GeV}, \\ \nonumber
  600~{\rm GeV} < \mttbar \leq 800~{\rm GeV}, \quad 800~{\rm GeV} < \mttbar \, .
 \end{eqnarray}
 For each of the four $\mttbar$ bins we select four bins in $y_p=\cos\theta_t^*$:
 \begin{equation} \label{eq:ybin}
  -1 \leq y_p < -\frac{1}{2}, \quad -\frac{1}{2} \leq y_p < 0, \quad 0 \leq y_p < \frac{1}{2},
   \quad \frac{1}{2} \leq y_p \leq 1 \, .
 \end{equation}
 
Our SM 	computations are performed at NLO QCD including the weak-interaction corrections. We refer to it with the acronym NLOW.
 In the calculation of the charge asymmetry $A_C$  also the mixed QCD-QED corrections of order $\alpha_s^2\alpha$ \cite{Bernreuther:2012sx}
 are taken into account in addition.\footnote{The LHC charge asymmetry $A_C$ was computed, for different c.m. energies, also  at NLO 
  and NNLO QCD in \cite{Kuhn:2011ri} and \cite{Czakon:2017lgo},
 respectively. In both references the electroweak corrections were taken into account, too. The relevance of the EW corrections
 for the charge asymmetry was first shown by \cite{Hollik:2011ps}.}

       The charge asymmetry and the  polarization and spin correlation observables $B, C$ used in this paper  are ratios. 
 They are, in   the SM at NLOW and to first order in the anomalous couplings
 schematically of the form
\begin{equation} \label{eq:BCratio}
 X = \frac{N_0+ N_1+ \delta N_{\rm NP}}{\sigma_0+  \sigma_1 + \delta\sigma_{\rm NP}} \, , \quad X= A_C, B, C. 
 \end{equation}
where $N_0$ $( N_1)$ and $\sigma_0$ $(\sigma_1)$ are the contributions  at LO QCD (NLOW) and $\delta N_{\rm NP}$
and $\delta\sigma_{\rm NP}$ denote  the first-order anomalous contributions to the numerator of the respective observables and the $t{\bar t}$ cross
section, respectively. We use this schematic notation both  for results inclusive in phase space and for bins in $\mttbar$ and $\cos\theta^*_t$.
 A priori, it is not clear how to evaluate these ratios where the numerator and denominator consist of truncated perturbation series.
 A typical Monte Carlo analysis would determine the numerators and denominators of \eqref{eq:BCratio} to the attainable order and evaluate the
 ratio without expanding it. In the spirit of perturbation theory one may expand the ratio. One gets at NLOW and to first order in the anomalous 
 couplings:
\begin{equation} \label{eq:Cexp}
X  = \left(1-\frac{\sigma_1}{\sigma_0}  - \frac{\delta \sigma_{\rm NP}}{\sigma_0} \right)\frac{N_0}{\sigma_0} +
\frac{N_1}{\sigma_0}  + \frac{\delta N_{\rm NP}}{\sigma_0}    + \mathcal{O}(\delta^2) \, .
\end{equation}
 The difference in the two prescriptions  \eqref{eq:BCratio} and \eqref{eq:Cexp}  for computing $X$
  are nominally of higher order in the SM and NP couplings. It may be considered as an additional theory uncertainty. 
  
 The results of our inclusive calculations  will  be given in expanded form. When computing our observables for the two-dimensional bins
 \eqref{mttbin}, \eqref{eq:ybin} all the six quantities in the ratio \eqref{eq:BCratio} will be separately determined. This allows for evaluation of the ratios
 in either way.  From the binned results listed in Appendix~\ref{sec:AppA} one can also obtain the inclusive results for 
 these quantities which allows for an unexpanded 
  evaluation of the ratios. Moreover, we display  our results for the three scale choices $\mu= m_t/2, m_t, 2 m_t$.  
  This allows to correctly account for the correlations of theory uncertainties when different observables of Table~\ref{tab:C-B-coeff} are combined, 
   which is advantageous for measurements (cf., e.g.  \cite{ATLAS:2016bac}).
  
  Table~\ref{tab:sigmAClowc} contains the $\ttbar$ cross section at NLOW for the three scales $\mu$. Theory predictions for 
   the cross section are, as is well known, 
  available at NNLO QCD \cite{Czakon:2013goa,Catani:2019iny}, including EW corrections \cite{Czakon:2017wor}. We need the NLOW result for the normalization of our observables. There are three contributions
  to $\stt$ from the NP Lagrangian \eqref{eq:LNPsum}. The effect of the chromomagnetic dipole operator is most significant and it is dominated by
  the contribution to $\ttbar$ production by $gg$ fusion. The contributions from the four-quark operators ${\cal O}_{VV}$ and ${\cal O}_3^1$ 
  are subdominant. Notice that the contributions from ${\cal O}_3^1$ to $u{\bar u}\to \ttbar$ and  $d{\bar d}\to \ttbar$  have opposite sign and 
  thus tend to cancel.\footnote{This remark applies also to the contributions of ${\cal O}_3^2$  and ${\cal O}_3^3$ to the other observables.}
  
  Table~\ref{tab:sigmAClowc} contains also the charge asymmetry $A_C$ in expanded form \eqref{eq:Cexp}. We recall that there is no contribution
  to its numerator at LO QCD. The NP contributions result from the  isospin-zero operator  ${\mathcal O}_{AA}$ and the P-even part of
  the isospin-one operator  ${\mathcal O}^1_2$. These operators induce contributions to the differential cross section that are odd
   under interchange of the $t$ and $\bar t$ momenta while those of the initial (anti)quark are kept fixed.

	\begin{table}[!h]
		\begin{center}
		\caption{ \label{tab:sigmAClowc} The SM and NP contributions to the $\ttbar$ cross 
		section and the LHC charge asymmetry \eqref{eq:LHCchaas} at $\sqrt{s}_{\rm had}=13.6$~TeV for three renormalization
		and factorization scales $\mu$.}
		  \vspace{1mm}
        {\renewcommand{\arraystretch}{1.2}
         \renewcommand{\tabcolsep}{0.2cm}
		\begin{tabular}{ c c c c c c }\hline
		\multicolumn{2}{c }{} 			& NLOW 	& $\propto \hat c_{VV}$ & $\propto \hat c_1$ & $\propto \Hmu$ \\ \hline\hline
		\multirow{3}{*}{$\sigma$[pb]} 	& $\mu = m_t/2$  &   874.71	& 826.06 &  97.00 &   3655.54             \\
						& $\mu = m_t$ 	 & 784.87      &  666.66 & 79.30 &    2838.32             \\
						& $\mu = 2m_t$	 &  693.55    &  546.86 &  65.78 &  2242.91     \\ \hline
		\multicolumn{2}{c}{} 	                & NLO + EW      & $\propto \hat c_{AA}$  & \multicolumn{2}{c }{$\propto \hat c_{2}$} \\ \hline\hline
		\multirow{3}{*}{$A_C$}		& $\mu = m_t/2$ & $7.41\times 10^{-3}$  &  0.335 & \multicolumn{2}{c}{ $6.87\times 10^{-2}$}  \\
						&  $\mu = m_t$	& $6.96\times 10^{-3}$  &  0.346 &  \multicolumn{2}{c}{$7.14\times 10^{-2}$} \\
						& $\mu = 2m_t$ 	& $6.48\times 10^{-3}$  &  0.357 & \multicolumn{2}{c}{$7.40\times 10^{-2}$}  \\   \hline
		
		\end{tabular}}
		\end{center}
	\end{table}%

Tables~\ref{tab:Ccoefnum} and~\ref{Bcoefnum} 	contain our results in expanded form for the spin correlations and polarizations at NLOW in the SM and the 
NP contributions from the effective Lagrangian \eqref{eq:LNPsum}.  
The dominant NP contribution to the first four spin correlations in table~\ref{tab:Ccoefnum}, which are P- and CP-even, is  from
the chromo-dipole operator that affects also $gg\to \ttbar$ besides $q{\bar q}\to\ttbar$. In particular, the correlation $C(r,r)$ which is small in the SM 
appears to have a good sensitivity to $\Hmu$.

The observables $C(k,k^*)$ and $C(r^*,k)+C(k,r^*)$ are the spin correlation analogues of $A_C$. The effect of using the  vectors $k^*$ and $r^*$ 
is that these correlations project onto different $y_p$ intervals than their un-starred analogues. They are sensitive to the couplings $\hcaa$
and $\xy2$. Therefore, they are useful, together with $A_C$, to obtain information about these couplings from experimental data, once they 
are available.

Likewise, the use of the vectors $r, k$ and $r^*, k^*$ in the polarization observables $B$ play analogous roles. The respective variables 
 project onto different $y_p$ intervals and are thus sensitive to different (combinations of) four-quark operators, 
  as the results
  of Table~\ref{Bcoefnum} show.
  
  The observable $B_1(n) + B_2(n)$ corresponds to the sum of the $t$ and $\bar t$ polarizations normal to the scattering plane
   and is generated by QCD absorptive parts \cite{Bernreuther:1995cx,Dharmaratna:1996xd,Bernreuther:2013aga}. The absorptive parts of the electroweak corrections to the $\ttbar$ production
   matrix elements contribute also, but are not shown here. They are roughly half of the QCD contributions and have the same sign. There are no contributions
   from the hermitean effective NP Lagrangian to LO QCD.
   
   The charge asymmetry $A_C$ and the spin correlation and  polarization observables of Tables~\ref{tab:Ccoefnum} 
 and~\ref{Bcoefnum}	provide a set that is large enough to measure, respectively constrain, the
  couplings of the effective NP Lagrangian \eqref{eq:LNPsum}.
  
  One may expect that the sensitivity to a specific anomalous coupling is not uniform in phase space. Therefore, 
we compute our observables also more differentially, namely, within the two-dimensional $(\mttbar, y_p)$ bins 
specified in \eqref{mttbin}, \eqref{eq:ybin} in order to investigate which region in phase space provides the highest 
sensitivity to a specific NP coupling. 
		
  One may ask whether any of these observables will depend, within a $(\mttbar, y_p)$ bin,  on additional NP parameters besides those 
 shown in Tables~\ref{tab:sigmAClowc} - \ref{Bcoefnum}.
		For instance, none of the four $y_p$ bins \eqref{eq:ybin} is  parity-symmetric; thus it could be 
		 that  the $P$-even observables have additional NP-parameter dependencies within a bin that cancel
		 in the sum over bins. We checked for all our observables  that within the  above $(\mttbar, y_p)$ bins 
		   no significant additional parameter dependencies appear; that is to say 
		 the numerical dependence on an additional NP parameter is at least 3 orders of magnitude smaller
         than the significant dependencies displayed in the tables of the appendix and are therefore discarded.

 	\begin{table}[!h]
				\begin{center}
		\caption{\label{tab:Ccoefnum}   The spin correlations $C$  at NLOW in the SM and the non-zero contributions
          of the NP Lagrangian \eqref{eq:LNPsum} for the c.m. energy $\sqrt{s}_{\rm had}=13.6$~TeV and 
          three renormalization and factorization scales $\mu$.}
           \vspace{1mm}
  {\renewcommand{\arraystretch}{1.0}
  \renewcommand{\tabcolsep}{0.2cm} 
		\begin{tabular}{c c c c c c }\hline
		\multicolumn{2}{ c }{}		      &  NLOW & $\propto \hat c_{VV}$ & $\propto \hat c_1$ & $\propto\Hmu$ \\\hline\hline	
		\multirow{3}{*}{$C(n,\,n)$} 	&$\mu=m_t/2$   & 0.324  & $-6.87\times 10^{-2}$ & 	 $-6.32 \times 10^{-3}$ &  2.061 \\
						&$\mu=m_t$   &	0.325	& $-7.73\times 10^{-2}$  & 	$-7.38\times 10^{-3}$&   2.037            \\
						& $\mu=2 m_t$  & 0.327		& $-8.53 \times 10^{-2}$ & $-8.36\times 10^{-3}$ & 2.014 \\ \hline
		\multirow{3}{*}{$C(r,\,r)$} 	& $\mu=m_t/2$ & $7.94\times 10^{-2}$   & $-0.676$ & $-7.88\times 10^{-2}$ 	& 2.506  \\
						& $\mu=m_t$ & 	$7.04\times 10^{-2}$	& $-0.703$		& $-8.30\times 10^{-2}$ & 2.487\\
						& $\mu=2 m_t$  &  $6.41\times 10^{-2}$		& $-0.728$ & $-8.70\times 10^{-2}$ & 2.468 \\     \hline
		\multirow{3}{*}{$C(k,\,k)$} 	&$\mu=m_t/2$   &  0.330	& $-1.195$ & $-0.143$ &  0.912 \\
						&$\mu=m_t$ & 0.331	 & $-1.234$ & $-0.149$ &  0.918 \\
						& $\mu=2 m_t$  & 0.333  & $-1.272$ & $-0.155$ &  0.924 \\ \hline
	\multirow{3}{*}{$C(r,\,k)+C(k,\,r)$}  &$\mu=m_t/2$  & $-0.203$ & $-0.294$ & $-3.25\times 10^{-2}$ & 0.736 \\
						                  &$\mu=m_t$&  $-0.206$  & $-0.308$ & $-3.45 \times 10^{-2}$ & 0.739 \\
						&$\mu=2 m_t$ & $-0.208$ & $-0.320$ & $-3.64\times 10^{-2}$ & 0.740    \\\hline
		\multicolumn{2}{ c }{}  & NLOW  & $\propto \hat c_{AA}$  & \multicolumn{2}{ c }{$\propto \hat c_{2}$} \\ \hline \hline
		\multirow{3}{*}{\scriptsize{$C(k,\,k^*)$}}&  $\mu=m_t/2$ &  $1.7\times 10^{-4}$   &  $-0.335$   & \multicolumn{2}{c}{ $-6.87\times 10^{-2}$} \\
							 & $\mu= m_t$ 	&   $2.1\times 10^{-4}$   &  $-0.346$  &  \multicolumn{2}{c}{$-7.14\times 10^{-2}$ } \\
							& $\mu=2 m_t$ 	& $2.5\times 10^{-4}$  & $-0.357$ &   \multicolumn{2}{c}{ $-7.40 \times 10^{-2}$} \\ \hline
		\multirow{3}{*}{\scriptsize{$C(r^*,\,k)\!+\!C(k,\,r^*)$}} & $\mu= m_t/2$ & $< 10^{-4}$ &  $-0.269$ & \multicolumn{2}{c}{ $-5.45\times 10^{-2}$ } \\
							& $\mu=m_t$ 	& $1.4\times 10^{-4}$ & $-0.281$	& \multicolumn{2}{c}{ $-5.74\times 10^{-2}$ } 	\\
							& $\mu=2 m_t$ 	& $ 3.4\times 10^{-4}$   &  $-0.293$	& \multicolumn{2}{c}{  $-6.01\times 10^{-2}$  } 	\\ \hline
		\multicolumn{2}{c}{} 					& \multicolumn{4}{c}{$\propto \Hd$} \\\hline\hline
		\multirow{3}{*}{\scriptsize{$C(n,\,r)\!-\!C(r,\,n)$}}	& $\mu=m_t/2$   & \multicolumn{4}{c}{ $-4.226$ } \\
							& $\mu=m_t$ 		&        \multicolumn{4}{c}{$-4.172$  } \\
							& $\mu=2 m_t$ 		&        \multicolumn{4}{c}{  $-4.119$ } \\ \hline
		\multirow{3}{*}{\scriptsize{$C(n,\,k)\!-\!C(k,\,n)$}}	& $\mu=m_t/2$  & \multicolumn{4}{c}{$-0.801$  } \\
							& $\mu=m_t$ 		&        \multicolumn{4}{c}{ $-0.799$ } \\
							& $\mu=2 m_t$ 		&        \multicolumn{4}{c}{ $-0.794$ } \\ \hline
		\end{tabular}}
		\end{center}
	       \end{table}

\newpage

	\begin{table}[!h]
		\begin{center}
		\caption{\label{Bcoefnum} The polarizations $B$ at NLOW in the SM and the non-zero contributions of the NP Lagrangian \eqref{eq:LNPsum} 
          for the c.m. energy $\sqrt{s}_{\rm had}=13.6$~TeV and three renormalization and factorization scales $\mu$.}
            \vspace{1mm}
     {\renewcommand{\arraystretch}{1.2}
      \renewcommand{\tabcolsep}{0.2cm}
		\begin{tabular}{c c c c c c c c }\hline
		\multicolumn{2}{ c }{} 			& \multicolumn{2}{ c }{NLOW} 			& \multicolumn{2}{ c }{$\propto \hat c_{VA}$} 		& \multicolumn{2}{ c }{$\propto \hat c_{3}$}  \\\hline\hline
		\multirow{3}{*}{$B_1(r) + B_2(r)$} & $\mu=m_t/2$ & \multicolumn{2}{ c }{ $ 1.6\times 10^{-3}$} & \multicolumn{2}{ c }{ 0.203  } & \multicolumn{2}{c}{$2.35\times 10^{-2}$} \\
						    & $\mu=m_t$	& \multicolumn{2}{ c }{$ 3.3 \times 10^{-3}$} 	& \multicolumn{2}{c}{  0.211  } 	& \multicolumn{2}{c}{$  2.49 \times 10^{-2}$} \\
						    & $\mu=2 m_t$	& \multicolumn{2}{c}{$  5.8 \times 10^{-3}$} 	& \multicolumn{2}{c}{0.220 } 	& \multicolumn{2}{c}{$   2.62 \times 10^{-2}$} \\\hline
		\multirow{3}{*}{$B_1(k) + B_2(k)$} &  $\mu=m_t/2$	& \multicolumn{2}{c}{$ 5.8 \times 10^{-3}  $} 	& \multicolumn{2}{c}{1.580} 	& \multicolumn{2}{c}{0.191} \\
						    & $\mu=m_t$	& \multicolumn{2}{c}{$ 8.4 \times 10^{-3}$} 	& \multicolumn{2}{c}{  1.628} 	& \multicolumn{2}{c}{ 0.199 } \\
						    & 	$\mu=2 m_t$& \multicolumn{2}{c}{$  1.21 \times 10^{-2}$} 	& \multicolumn{2}{c}{ 1.678} 	& \multicolumn{2}{c}{0.208 } \\ \hline
		\multicolumn{2}{c}{} & \multicolumn{2}{c}{NLOW }& \multicolumn{2}{c}{$\propto \hat c_{AV}$} & \multicolumn{2}{c}{$\propto \hat c_{1} - \hat c_{2} + \hat c_{3}$}   \\\hline\hline
		\multirow{3}{*}{$B_1(r^*) + B_2(r^*)$}	 & $\mu=m_t/2$ &\multicolumn{2}{c}{$< 10^{-4}$} 	& \multicolumn{2}{c}{0.752 } 	& \multicolumn{2}{c}{ 0.152 }   \\
							& $\mu=m_t$& \multicolumn{2}{c}{$1.7 \times 10^{-4} $} 	& \multicolumn{2}{c}{ 0.790} 	& \multicolumn{2}{c}{ 0.161 } \\
							& $\mu=2 m_t$ & \multicolumn{2}{c}{$8.3 \times 10^{-4} $} 	& \multicolumn{2}{c}{  0.828} 	& \multicolumn{2}{c}{  0.169} \\\hline
		\multirow{3}{*}{$B_1(k^*) + B_2(k^*)$}  &$\mu=m_t/2$ & \multicolumn{2}{c}{$ < 10^{-4} $} 	& \multicolumn{2}{c}{0.858} 	& \multicolumn{2}{c}{0.175 }  \\
							& $\mu=m_t$& \multicolumn{2}{c}{$ 4.9\times  10^{-4}   $} 	& \multicolumn{2}{c}{  0.891} 	& \multicolumn{2}{c}{ 0.183 } \\
							& $\mu=2 m_t$ & \multicolumn{2}{c}{$ 1.3\times 10^{-3}  $} 	& \multicolumn{2}{c}{0.925 } 	& \multicolumn{2}{c}{  0.191} \\\hline
		\multicolumn{2}{c}{} &    \multicolumn{6}{c}{NLO QCD}  \\\hline \hline
		\multirow{3}{*}{$B_1(n) + B_2(n)$}	& $\mu=m_t/2$ &  \multicolumn{6}{c}{$ 8.40 \times 10^{-3}$} \\
							& $\mu=m_t$   &  \multicolumn{6}{c}{$ 7.50\times 10^{-3}$} \\
							&$\mu=2 m_t$  &  \multicolumn{6}{c}{$  6.77\times 10^{-3}$} \\\hline
		
		\end{tabular} }
		\end{center}
		\end{table}

  \begin{table}[!htb]
\caption{As in Table~\ref{tab:sigmAClowc}  but for the two-dimensional bin $M_{t\bar t}>600$ GeV and $-0.5\leq y_p\leq 0.5$.
		 \label{tab:sighbin}}
		  \vspace{1mm}
		\begin{center}  
        {\renewcommand{\arraystretch}{1.2}
         \renewcommand{\tabcolsep}{0.2cm}
		\begin{tabular}{ c c c c c c }\hline
\multicolumn{2}{c }{} 	& NLOW 	& $\propto \hat c_{VV}$ & $\propto \hat c_1$ & $\propto \Hmu$ \\ \hline\hline		
\multirow{3}{*}{$\sigma$(pb)} & $\mu=m_{t}/2$ & $30.04$ & 168.29    &  21.34   & 233.56 \tabularnewline 
                              & $\mu=m_{t}$ & $31.00$   & 132.52    &  17.05   & 176.47\tabularnewline
                               & $\mu=2m_{t}$ & $29.44$ & 106.36      &  13.85   &  136.26  \tabularnewline \hline 
\multicolumn{2}{c}{}       & NLO + EW & $\propto \hat c_{AA}$  & \multicolumn{2}{c }{$\propto \hat c_{2}$} \\ \hline\hline
 \multirow{3}{*}{$A_{C}$} & $\mu=m_{t}/2$ & $7.03\times10^{-3}$&  0.885  &\multicolumn{2}{c}{ 0.185 }  \tabularnewline
                          & $\mu=m_{t}$ & $6.74\times10^{-3}$&  0.920  &\multicolumn{2}{c}{ 0.193 }\tabularnewline
                          & $\mu=2m_{t}$ & $6.55\times10^{-3}$&  0.955  &\multicolumn{2}{c}{ 0.202}\tabularnewline \hline
\end{tabular} }
\end{center}
\end{table}
 	\begin{table}[!htb]
				\begin{center}
		\caption{As in Table~\ref{tab:Ccoefnum}   but for the two-dimensional bin $M_{t\bar t}>600$ GeV and $-0.5\leq y_p\leq 0.5$.
          \label{tab:Chbin}}
           \vspace{1mm}
  {\renewcommand{\arraystretch}{1.0}
  \renewcommand{\tabcolsep}{0.2cm} 
		\begin{tabular}{c c c c c c }\hline
		\multicolumn{2}{ c }{}		      &  NLOW & $\propto \hat c_{VV}$ & $\propto \hat c_1$ & $\propto\Hmu$ \\\hline\hline	
		\multirow{3}{*}{$C(n,\,n)$} 	&$\mu=m_t/2$   & 0.414  & 1.176 &  0.547 &  2.101 \\
						&$\mu=m_t$   &	0.418	& 1.193 & 0.551 &   2.037            \\
						& $\mu=2 m_t$  & 0.421		& 1.211 & 0.555 & 1.975\\ \hline
		\multirow{3}{*}{$C(r,\,r)$} 	& $\mu=m_t/2$ & -0.530   & -1.673 & -0.745 &  2.949  \\
						                & $\mu=m_t$ & 	-0.539	& -1.710 &-0.754  & 2.889 \\
						                & $\mu=2 m_t$  & -0.547	& -1.745 & -0.762 & 2.831 \\     \hline
		\multirow{3}{*}{$C(k,\,k)$} 	&$\mu=m_t/2$   &  -0.254 & -2.315 & -0.534 &  2.014 \\
						                &$\mu=m_t$     & -0.255	 & -2.379 & -0.548  &  2.017 \\
						                & $\mu=2 m_t$  & -0.257  & -2.440 & -0.563 &  2.021 \\ \hline
	\multirow{3}{*}{$C(r,\,k)+C(k,\,r)$} &$\mu=m_t/2$  & -0.230 & -0.388 & -0.273 & -0.945\\
						                & $\mu=m_t$&  -0.234 & -0.397 & -0.276 & -0.944 \\
						                 &$\mu=2 m_t$ & -0.238 & -0.406 & -0.278 & -0.946   \\\hline
		\multicolumn{2}{ c }{}  & NLOW  & $\propto \hat c_{AA}$  & \multicolumn{2}{ c }{$\propto \hat c_{2}$} \\ \hline \hline
		\multirow{3}{*}{\scriptsize{$C(k,\,k^*)$}}&  $\mu=m_t/2$ &  $0.5\times 10^{-4}$   &  -0.885   & \multicolumn{2}{c}{ -0.185} \\
							 & $\mu= m_t$ 	&   $3.0\times 10^{-4}$   &  -0.920 &  \multicolumn{2}{c}{-0.193} \\
							& $\mu=2 m_t$ 	& $3.8\times 10^{-4}$  & -0.955 &   \multicolumn{2}{c}{ -0.202} \\ \hline
		\multirow{3}{*}{\scriptsize{$C(r^*,\,k)\!+\!C(k,\,r^*)$}} & $\mu= m_t/2$ & $3.7\times 10^{-4}$ &  -1.278 & \multicolumn{2}{c}{-0.265} \\
							& $\mu=m_t$ 	& $3.9 \times 10^{-4}$ & -1.341	& \multicolumn{2}{c}{ -0.280 } 	\\
							& $\mu=2 m_t$ 	& $ 2.1 \times 10^{-4}$   &  -1.402	& \multicolumn{2}{c}{ -0.294  } 	\\ \hline
		\multicolumn{2}{c}{} 					& \multicolumn{4}{c}{$\propto \Hd$} \\\hline\hline
		\multirow{3}{*}{\scriptsize{$C(n,\,r)\!-\!C(r,\,n)$}}	& $\mu=m_t/2$   & \multicolumn{4}{c}{ $-4.524$ } \\
							& $\mu=m_t$ 		&        \multicolumn{4}{c}{$-4.395$  } \\
							& $\mu=2 m_t$ 		&        \multicolumn{4}{c}{  $-4.270$ } \\ \hline
		\multirow{3}{*}{\scriptsize{$C(n,\,k)\!-\!C(k,\,n)$}}	& $\mu=m_t/2$  & \multicolumn{4}{c}{1.261 } \\
							& $\mu=m_t$ 		&        \multicolumn{4}{c}{1.263 } \\
							& $\mu=2 m_t$ 		&        \multicolumn{4}{c}{ 1.269 } \\ \hline
		\end{tabular}}
		\end{center}
	       \end{table}


	\begin{table}[!htb]
		\begin{center}
		\caption{As in Table~\ref{Bcoefnum}  but for the two-dimensional bin $M_{t\bar t}>600$ GeV and $-0.5\leq y_p\leq 0.5$.  
		\label{tab:Bhbin}}
            \vspace{1mm}
     {\renewcommand{\arraystretch}{1.2}
      \renewcommand{\tabcolsep}{0.2cm}
		\begin{tabular}{c c c c c c c c }\hline
		\multicolumn{2}{ c }{} 			& \multicolumn{2}{ c }{NLOW} 			& \multicolumn{2}{ c }{$\propto \hat c_{VA}$} 		& \multicolumn{2}{ c }{$\propto \hat c_{3}$}  \\\hline\hline
		\multirow{3}{*}{$B_1(r) + B_2(r)$} & $\mu=m_t/2$ 	& \multicolumn{2}{ c }{ $ 4.1\times 10^{-3}$} 	& \multicolumn{2}{ c }{ 0.547 } 	& \multicolumn{2}{c}{$6.80\times  10^{-2}$} \\
						    & $\mu=m_t$	& \multicolumn{2}{ c }{$ 4.5\times 10^{-3}$} 	& \multicolumn{2}{c}{  0.573  } 	& \multicolumn{2}{c}{$  7.24 \times 10^{-2}$} \\
						    & $\mu=2 m_t$	& \multicolumn{2}{c}{$  5.0 \times 10^{-3}$} 	& \multicolumn{2}{c}{0.599} 	& \multicolumn{2}{c}{$   7.66 \times 10^{-2}$} \\\hline
		\multirow{3}{*}{$B_1(k) + B_2(k)$} &  $\mu=m_t/2$	& \multicolumn{2}{c}{0.013} 	& \multicolumn{2}{c}{6.717} 	& \multicolumn{2}{c}{0.860} \\
						    & $\mu=m_t$	& \multicolumn{2}{c}{0.012} 	& \multicolumn{2}{c}{  6.980} 	& \multicolumn{2}{c}{ 0.907 } \\
						    & 	$\mu=2 m_t$& \multicolumn{2}{c}{0.011} 	& \multicolumn{2}{c}{ 7.237} 	& \multicolumn{2}{c}{0.952 } \\ \hline
		\multicolumn{2}{c}{} & \multicolumn{2}{c}{NLOW }& \multicolumn{2}{c}{$\propto \hat c_{AV}$} & \multicolumn{2}{c}{$\propto \hat c_{1} - \hat c_{2} + \hat c_{3}$}   \\\hline\hline
		\multirow{3}{*}{$B_1(r^*) + B_2(r^*)$}	 & $\mu=m_t/2$ &\multicolumn{2}{c}{$7.7\times 10^{-4}$} 	& \multicolumn{2}{c}{2.842 } 	& \multicolumn{2}{c}{ 0.589 }   \\
							& $\mu=m_t$& \multicolumn{2}{c}{$1.1 \times 10^{-3} $} 	& \multicolumn{2}{c}{ 2.987} 	& \multicolumn{2}{c}{ 0.623 } \\
							& $\mu=2 m_t$ & \multicolumn{2}{c}{$1.3 \times 10^{-3} $} 	& \multicolumn{2}{c}{  3.127} 	& \multicolumn{2}{c}{  0.655} \\\hline
		\multirow{3}{*}{$B_1(k^*) + B_2(k^*)$}  &$\mu=m_t/2$ & \multicolumn{2}{c}{$7.5 \times 10^{-4} $} 	& \multicolumn{2}{c}{1.935} 	& \multicolumn{2}{c}{0.403}  \\
							& $\mu=m_t$& \multicolumn{2}{c}{$ 7.9 \times 10^{-4}   $} 	& \multicolumn{2}{c}{  2.016} 	& \multicolumn{2}{c}{ 0.423 } \\
							& $\mu=2 m_t$ & \multicolumn{2}{c}{$ 8.5 \times 10^{-4}  $} 	& \multicolumn{2}{c}{2.094 } 	& \multicolumn{2}{c}{  0.442} \\\hline
		\multicolumn{2}{c}{} &    \multicolumn{6}{c}{NLO QCD}  \\\hline \hline
		\multirow{3}{*}{$B_1(n) + B_2(n)$}	& $\mu=m_t/2$ &  \multicolumn{6}{c}{$1.86 \times 10^{-2}$} \\
							& $\mu=m_t$   &  \multicolumn{6}{c}{$ 1.67\times 10^{-2}$} \\
							&$\mu=2 m_t$  &  \multicolumn{6}{c}{$ 1.51\times 10^{-2}$} \\\hline
		
		\end{tabular} }
		\end{center}
		\end{table}


Our results for the binned $\stt$, $A_C$, and the spin observables are given in Tables~\ref{tab:binsig} - \ref{tab:binNP42m}
 of appendix~\ref{sec:AppA}. As already mentioned above,  we compute for each observable each of the six quantities (if non-zero) 
 in the ratio \eqref{eq:BCratio} separately. This allows to compute $A_C$ and the spin observables either in unexpanded
  or expanded form.
  
An inspection of the bins in Tables~\ref{tab:binNP1hm} - \ref{tab:binNP42m}  of the NP contributions to our observables indicates that 
in almost all cases the two bins at large $\ttbar$ invariant mass $\mttbar>800$~GeV in the central region, $-0.5\leq y_p\leq 0.5$
seem to have the highest sensitivity to the parameters of the effective NP Lagrangian. In order to eventually obtain a reasonable large
dileptonic $\ttbar$ data sample, we suggest here to consider the phase-space region $\mttbar>600$~GeV and $-0.5\leq y_p\leq 0.5$.
Assuming that at the LHC at 13.6~TeV an integrated luminosity of 300 ${\rm fb}^{-1}$ will eventually
  be collected, and using the $\ttbar$ cross section given in Table~\ref{tab:sighbin} for an estimate, one expects about $5\times 10^5$ dileptonic $\ell\ell'$ $(\ell,\ell'=e,\mu)$ events in this region.

We use the four two-dimensional bins $\mttbar>600$~GeV and $-0.5\leq y_p\leq 0.5$ of 
  Tables~\ref{tab:binsig} - \ref{tab:binNP42m} and compute, by summing the respective numbers of the four bins, the charge asymmetry and the normalized
   spin observables in this phase-space region in expanded form \eqref{eq:Cexp}. The results are given Tables~\ref{tab:sighbin} - \ref{tab:Bhbin}.
  Comparing the coefficients that multiply the contributions of the NP parameters ${\hat c}_{IJ}, (I,J=V,A)$, $\xx1,  \xy2,\xz3$ 
  in these tables with the respective numbers in Tables~\ref{tab:sigmAClowc} - \ref{Bcoefnum} of the inclusive results, one sees that 
  the sensitivity to the couplings of the four-quark operators increases significantly in the high $\mttbar$, central region. 
  In particular, the spin correlations $C(k, k^*)$ and $C(r^*, k) + C(k, r^*)$ appear to be useful for disentangling the contributions 
   from the operators associated with $\hcaa$ and $\xy2$. Not much is gained in this phase-space region 
    for the sensitivity to the chromo-dipole moments $\Hmu$ and $\Hd$ compared with the inclusive case.

\section{Summary }
 \label{sec:sumconc}		
 We have elaborated on a set of spin correlation and polarization observables, proposed previously by two of 
 the authors of this paper,
 that allows to probe the hadronic $\ttbar$ production dynamics in detail. These observables project 
 out all entries of the hadronic production spin density 
 matrices. We considered $\ttbar$ production and decays into dileptonic final states at the LHC for the present c.m. energy 13.6~TeV. We computed these
 observables within the Standard Model at NLO QCD including the mixed QCD weak-interaction contributions. Possible new physics effects were incorporated 
 by using an  $SU(3)_c \times SU(2)_L \times U(1)_{Y}$ effective Lagrangian with operators that generate  tree-level interferences with the LO QCD $gg, \qqbar \to \ttbar$ 
  amplitudes. The effect of these NP operators on our observables  was taken into account to linear order in the anomalous couplings. This can be justified 
  by the results of the CMS experiment at 13.6 TeV \cite{CMS:2019nrx} which show that these dimensionless couplings must be markedly smaller than one. 
  We considered also the LHC charge asymmetry $A_C$ and two additional spin correlations that turn out to be very useful in disentangling 
  contributions from NP four-quark operators. We emphasize that several our of observables allow for direct searches of non-standard P and CP violation 
   in $\ttbar$ events.
  
  In addition to  computing our observables inclusive in phase space, we determined them also in two-dimensional 
  $(\mttbar,\cos\theta_t^*)$ bins, where $\mttbar$ denotes the $\ttbar$ invariant mass  and
$\theta_t^*$ is the top-quark scattering angle in the $\ttbar$ zero-momentum frame. 
Our analysis shows that the sensitivity to a number of anomalous couplings
 significantly increases in the high-energy central region.
 
 Experimental measurements of these observables were so far made only inclusively, and no deviation from the SM was found. 
  The contributions of anomalous couplings to an observable are, however, not uniform in phase space as our results show.
   More differential measurements in the future, especially in the  high-energy central region, promise to significantly increase our 
   knowledge about top-quark interactions beyond the Standard Model.
 

\section*{Acknowledgments}
We thank A. Grohsjean, Ch. Schwanenberger, and A. Zimermmane-Santos for discussions.
The work of Z.G.S. and L.C. were supported by the Natural Science Foundation of China under 
contract No.~12235008 and No.~12321005. The work of L.C. was supported also by the Natural Science Foundation 
of China under contract No.~1220517, and by the Taishan Scholar Foundation of Shandong province (tsqn202312052).


\pagebreak
\appendix
\section{SM and NP values of the binned observables}
\label{sec:AppA}

In the tables of this appendix, we present our results for the cross section $\stt$, the charge asymmetry 
$A_C$, and the spin correlations and polarization observables defined in Section~\ref{sec:obs} and 
listed in Table~\ref{tab:C-B-coeff} for the two-dimensional bins
 \eqref{mttbin}, \eqref{eq:ybin} at $\sqrt{s_{\rm had}}=13.6$~TeV. For the cross section, we list the
 respective values at LO QCD, the contributions at NLOW, and to first order in the NP couplings.
 For $A_C$ and the spin observables, we list the  value $N_0$ of the respective numerator (cf. \eqref{eq:BCratio})
 -- if it is significantly different from zero --, the contributions $N_1$ at NLOW, and those of the NP operators $\delta N_{NP}$.
 These quantities allow to compute $A_C$ and the spin observables either in unexpanded or expanded form, cf.
  eqs.~\eqref{eq:BCratio} and \eqref{eq:Cexp}. \\
Tables~\ref{tab:binsig} - \ref{tab:binBabs} contain our SM results for the  three chosen scales $\mu$, 
while Tables~\ref{tab:binNP1hm} - \ref{tab:binNP42m}
 contain the various NP contributions to each observable. Here the values for the three scales are shown in separate tables.

		\begin{table}[htb]
		\begin{center}
		\caption{ \label{tab:binsig} The $\ttbar$ cross section at 13.6~TeV in bins of $\mttbar$ and $y_p=\cos\theta_t^*$ for 3 scales $\mu$.
		For each invariant mass bin the first column displays the range of the $y_p$ bin. The numbers in the 2nd, 3rd, and 4th column are 
		the values of $\sigma_{\ttbar}$ at LO QCD $(\sigma_0)$ for $\mu=m_t/2, m_t,$ and $2m_t$, respectively. The 5th, 6th, and 7th column contain the
		NLO QCD plus weak-interaction  contributions to $\sigma_{\ttbar}$ $(\sigma_1)$ for $\mu=m_t/2, m_t,$ and $2m_t$, respectively.
		All cross-section numbers are in units of pb.}	
	  \vspace{1mm}
  {\renewcommand{\arraystretch}{1.2}
  \renewcommand{\tabcolsep}{0.2cm}	
  \begin{tabular}{c c c c c c c }\hline	
\multicolumn{7}{ c }{ $ 2m_t\le \mttbar \le 450~{\rm GeV}$}  \\ \hline \hline
 (  -1.0,  -0.5) & 74.596   & 59.133  &  47.579  &  30.112   & 33.509   & 33.911 \\
 (  -0.5,   0.0)  & 58.987   & 46.915   & 37.852  &  22.347   & 25.519  &  26.202\\
 (   0.0,   0.5)  & 59.071  &  46.971  &  37.876  &  18.071   &  22.316  &  23.847\\
 (   0.5,   1.0)  & 74.355  &  58.987  &  47.488    & 29.831  &  33.313   & 33.780\\ \hline\hline
 \multicolumn{7}{ c }{ $ 450~{\rm GeV}<\mttbar\le 600~{\rm GeV} $}  \\ \hline \hline 
 (  -1.0,  -0.5)  & 81.987   & 63.546   & 50.087  &  28.030  &  33.922  &  35.307\\
 (  -0.5,   0.0)  & 42.980   & 33.454   & 26.488   &  2.410   &  9.103   & 12.139\\
 (   0.0,   0.5)  & 42.855   & 33.371  &  26.433   &  2.273   &  8.823   & 12.025\\
 (   0.5,   1.0)  & 82.015  &  63.521 &   50.054  &  28.667   & 34.385 &   35.554\\ \hline\hline
\multicolumn{7}{ c }{ $ 600~{\rm GeV} < \mttbar \le  800~{\rm GeV} $}  \\ \hline \hline 
 (  -1.0,  -0.5)  & 42.366  &  31.993   & 24.665   & 12.792   & 16.603 &   17.532\\
 (  -0.5,   0.0)  & 14.643    & 11.144  &   8.667   & -2.538   &  0.910   &  2.622\\
 (   0.0,   0.5)  & 14.653  &  11.151    & 8.662   & -2.118   &  1.234   &  2.858\\
 (   0.5,   1.0)  & 42.413   & 32.029  &  24.698  &  12.066  &  16.116  &  17.234\\ \hline\hline
 \multicolumn{7}{ c }{ $\mttbar > 800~{\rm GeV} $}  \\ \hline \hline 
 (  -1.0,  -0.5)   & 23.747   & 17.292   & 12.915   &  4.418   &  7.836   &  8.750\\
 (  -0.5,   0.0)   & 5.500   &  4.072   &  3.087   & -2.769   & -0.768   &  0.241\\
 (   0.0,   0.5)   & 5.499   &  4.069   &  3.084   & -2.832  &  -0.809   &  0.221\\
 (   0.5,   1.0)  & 23.753   & 17.301   & 12.924    & 4.526   &  7.905  &   8.762\\ \hline\hline
 \end{tabular}}
		\end{center}
	\end{table}%

		\begin{table}[h]
		\centering
		\caption{ \label{tab:binchaasy} The numerator of the LHC charge asymmetry $A_C$ in the SM
		 defined in Eq.~\eqref{eq:LHCchaas} at 13.6~TeV in bins of $\mttbar$ 
		 and $y_p=\cos\theta_t^*$ for 3 scales $\mu$.
		For each invariant mass bin the first column displays the range of the $y_p$ bin. The numbers in the 2nd, 3rd, and 4th column are 
		the values of $A_C$ at NLO QCD plus electroweak interactions for $\mu=m_t/2, m_t,$ and $2m_t$, respectively. 
		All numbers are in units of pb.}
				  \vspace{1mm}
  {\renewcommand{\arraystretch}{1.2}
    \renewcommand{\tabcolsep}{0.2cm}
 \begin{tabular}{c c c c }\hline	
\multicolumn{4}{ c }{ $ 2m_t\le \mttbar \le 450~{\rm GeV}$}  \\ \hline \hline
 (  -1.0,  -0.5) &  0.614  &   0.452  &    0.336 \\
 (  -0.5,   0.0) &   0.114  &   0.082 &    0.062\\
 (   0.0,   0.5) &   0.129  &   0.098   &  0.068\\
 (   0.5,   1.0)  &  0.584  &   0.428   &  0.317 \\ \hline \hline
\multicolumn{4}{ c }{ $ 450~{\rm GeV}<\mttbar\le 600~{\rm GeV} $}  \\ \hline \hline 
 (  -1.0,  -0.5) &   0.729   &  0.536  &   0.396\\
 (  -0.5,   0.0) &   0.140 &    0.116  &   0.093\\
 (   0.0,   0.5)  &  0.199   &  0.147  &   0.111 \\ 
 (   0.5,   1.0)  &  0.799  &   0.581  &   0.422 \\ \hline \hline
\multicolumn{4}{ c }{ $ 600~{\rm GeV} < \mttbar \le  800~{\rm GeV} $}  \\ \hline \hline 
 (  -1.0,  -0.5)  &  0.433   &  0.311 &    0.225\\
 (  -0.5,   0.0)  &  0.086  &   0.061   &  0.046\\
 (   0.0,   0.5)  &  0.089 &    0.065  &   0.052\\
 (   0.5,   1.0)  &  0.445 &    0.320  &   0.234 \\ \hline \hline
\multicolumn{4}{ c }{ $\mttbar > 800~{\rm GeV} $}  \\ \hline \hline 
 (  -1.0,  -0.5) &   0.321  &   0.224  &   0.160\\
 (  -0.5,   0.0)   & 0.054  &   0.039 &    0.028\\
 (   0.0,   0.5) &   0.054   &  0.040 &    0.028\\
 (   0.5,   1.0) &   0.320  &   0.225  &   0.159 \\ \hline \hline
\end{tabular}}
	\end{table}%

		\begin{table}[h]
		\centering
		\vspace{1mm}
  {\renewcommand{\arraystretch}{1.2}
    \renewcommand{\tabcolsep}{0.2cm}
    
		\caption{ \label{tab:binCnn} The numerator of the spin correlation $C(n,n) = N_{nn}/\sigma_{\ttbar}$ in the SM
		at 13.6~TeV in bins of $\mttbar$ and $y_p=\cos\theta_t^*$ for 3 scales $\mu$.
		For each invariant mass bin the first column displays the range of the $y_p$ bin. The numbers in the 2nd, 3rd, and 4th column are 
		the values of $N_{0,nn}$ at LO QCD for $\mu=m_t/2, m_t,$ and $2m_t$, respectively. The 5th, 6th, and 7th column contain the
		NLO QCD plus weak-interaction  contributions $N_{1,nn}$ for $\mu=m_t/2, m_t,$ and $2m_t$, respectively.
		All numbers are in units of pb.}			
		\vspace{1mm}
  \begin{tabular}{c c c c c c c }\hline	
\multicolumn{7}{c}{ $ 2m_t\le \mttbar \le 450~{\rm GeV}$}  \\ \hline \hline 
 (  -1.0,  -0.5) & 34.099   &  26.853  &   21.453  &   13.612  &   15.427   &  15.728 \\
 (  -0.5,   0.0)  &  25.677   &  20.277   &  16.243  & 8.558  &   10.361 &  10.918\\
 (   0.0,   0.5)  & 25.708    & 20.301    &  16.262 &   6.724 &   9.037  &  9.928\\
 (   0.5,   1.0) & 34.004   & 26.777 &  21.395  & 13.698 &  15.473 &  15.740\\ \hline \hline 
 \multicolumn{7}{c}{ $ 450~{\rm GeV}<\mttbar\le 600~{\rm GeV} $}  \\ \hline \hline 
 (  -1.0,  -0.5) & 19.913 &  15.398  & 12.107  &  8.343 &   9.303  &  9.339\\
 (  -0.5,   0.0) & 13.877  & 10.798  &  8.549   & 0.230 &   2.525 &   3.650\\
 (   0.0,   0.5)  & 13.833  & 10.765   & 8.527 &   0.208  &  2.491 &   3.588\\
 (   0.5,   1.0) & 19.913  & 15.396  & 12.104 &   8.491 &   9.408 &    9.432\\ \hline \hline 
\multicolumn{7}{c}{ $ 600~{\rm GeV}< \mttbar \le  800~{\rm GeV} $}  \\ \hline \hline
 (  -1.0,  -0.5)  & 6.998 &   5.291  &  4.084  &  2.371&    2.927 &    3.013\\
 (  -0.5,   0.0)  & 5.954  &  4.548  &  3.544  & -1.334  &  0.162  &  0.916\\
 (   0.0,   0.5) &  5.959   & 4.550  &  3.544  & -1.137 &   0.295 &   1.019\\
 (   0.5,   1.0)  & 7.005   & 5.296  &  4.089 &   2.191  &  2.807  &  2.926\\ \hline \hline 
\multicolumn{7}{c}{ $\mttbar > 800~{\rm GeV} $}  \\ \hline \hline 
 (  -1.0,  -0.5)  & 3.212  &  2.352  &  1.764  &  0.168   &  0.758   &  0.973\\
 (  -0.5,   0.0)   & 3.104  &   2.300  &   1.748 &   -1.784 &   -0.588 &  0.023\\
 (   0.0,   0.5) &   3.105   &  2.302  &   1.749 &   -1.817 &   -0.602 &   0.017\\
 (   0.5,   1.0) &   3.218  &   2.355  &   1.767  &   0.203  &   0.768   &  0.982\\ \hline \hline 
	\end{tabular}}
 
	\end{table}%

		\begin{table}[h]
		\centering
	\vspace{1mm}
  {\renewcommand{\arraystretch}{1.2}
    \renewcommand{\tabcolsep}{0.2cm}
		\caption{ \label{tab:binCrr} The numerator of the spin correlation $C(r,r) = N_{rr}/\sigma_{\ttbar}$ in the SM
		at 13.6~TeV in bins of $\mttbar$ and $y_p=\cos\theta_t^*$ for 3 scales $\mu$.
		For each invariant mass bin the first column displays the range of the $y_p$ bin. The numbers in the 2nd, 3rd, and 4th column are 
		the values of $N_{0,rr}$ at LO QCD for $\mu=m_t/2, m_t,$ and $2m_t$, respectively. The 5th, 6th, and 7th column contain the
		NLO QCD plus weak-interaction  contributions $N_{1,rr}$ for $\mu=m_t/2, m_t,$ and $2m_t$, respectively.
		All numbers are in units of pb.}	
		\vspace{1mm}
  \begin{tabular}{c c c c c c c }\hline	
\multicolumn{7}{ c }{ $ 2m_t\le \mttbar \le 450~{\rm GeV}$}  \\ \hline \hline 
 (  -1.0,  -0.5) & 23.907 &   18.608  &  14.684 &   11.339 &   12.228 &   12.187 \\
 (  -0.5,   0.0) &   0.491  &  -0.042 &   -0.382  &   3.378  &   2.773  &   2.259\\
 (   0.0,   0.5)  &  0.466  &  -0.062 &   -0.400 &    3.059 &    2.502 &    2.043\\
 (   0.5,   1.0) &  23.824 &   18.538 &   14.631  &  11.350 &   12.261 &   12.215\\ \hline \hline 
\multicolumn{7}{ c }{ $ 450~{\rm GeV} <\mttbar\le 600~{\rm GeV} $}  \\ \hline \hline 
 (  -1.0,  -0.5)   & 4.665   &  3.468  &   2.618 &    6.238 &    5.301 &    4.514\\
 (  -0.5,   0.0) & -16.485  & -13.003 &  -10.423  &   2.112 &   -1.292 &   -3.070\\
 (   0.0,   0.5) & -16.458  & -12.972 &  -10.407  &   2.108  &  -1.266 &   -3.054\\
 (   0.5,   1.0)  &  4.666  &   3.467  &   2.620  &   6.308  &   5.334 &   4.525\\ \hline \hline 
\multicolumn{7}{ c }{ $ 600~{\rm GeV} < \mttbar \le  800~{\rm GeV} $}  \\ \hline \hline 
 (  -1.0,  -0.5) & -2.359 & -1.824  & -1.438  &  1.678 &   0.739  &  0.216\\
 (  -0.5,   0.0) & -8.431 &  -6.459 &  -5.051  &  2.511 &   0.211  & -0.990\\
 (   0.0,   0.5) & -8.440  & -6.463 &  -5.054 &   2.287 &   0.046 &  -1.120\\
 (   0.5,   1.0) & -2.356  & -1.823  & -1.436  &  1.801  &  0.823  &  0.283\\ \hline \hline 
\multicolumn{7}{ c }{ $\mttbar > 800~{\rm GeV} $}  \\ \hline \hline 
 (  -1.0,  -0.5)  & -2.457  & -1.809 &  -1.364  &  0.997  &  0.171  & -0.227\\
 (  -0.5,   0.0)  & -3.866 &  -2.870  & -2.183  &  2.345 &   0.810  &  0.016\\
 (   0.0,   0.5)  & -3.863  & -2.870  & -2.182  &  2.372 &   0.838  &  0.043\\
 (   0.5,   1.0)  & -2.464  & -1.811  & -1.366  &  0.990  &  0.154 &  -0.237\\ \hline \hline 
\end{tabular}}
 
	\end{table}%

		\begin{table}[h]
		\centering
		  {\renewcommand{\arraystretch}{1.2}
                  \renewcommand{\tabcolsep}{0.2cm}
		\caption{ \label{tab:binCkk} The numerator  of the spin correlation $C(k,k) = N_{kk}/\sigma_{\ttbar}$ in the SM
		at 13.6~TeV in bins of $\mttbar$ and $y_p=\cos\theta_t^*$ for 3 scales $\mu$.
		For each invariant mass bin the first column displays the range of the $y_p$ bin. The numbers in the 2nd, 3rd, and 4th column are 
		the values of $N_{0,kk}$ at LO QCD for $\mu=m_t/2, m_t,$ and $2m_t$, respectively. The 5th, 6th, and 7th column contain the
		NLO QCD plus weak-interaction  contributions  $N_{1,kk}$ for $\mu=m_t/2, m_t,$ and $2m_t$, respectively.
		All numbers are in units of pb.}	
		\vspace{1mm}
  \begin{tabular}{ c c c c c c c }\hline	
\multicolumn{7}{ c }{ $ 2m_t\le \mttbar \le 450~{\rm GeV}$}  \\ \hline \hline  
 (  -1.0,  -0.5) & 40.027   &  31.090   &  24.493  &   11.974   &  15.708  &   17.026 \\
 (  -0.5,   0.0)  &  29.272   &  22.926  &   18.219   &   8.820  &   11.339   &  12.214\\
 (   0.0,   0.5)  &  29.295  &   22.939  &   18.229  &    6.653 &     9.746  &   10.986\\
 (   0.5,   1.0)   & 39.892  &   30.981  &   24.399  &   11.936  &   15.683  &   17.043  \\ \hline \hline 
\multicolumn{7}{ c }{ $ 450~{\rm GeV} <\mttbar\le 600~{\rm GeV} $}  \\ \hline \hline 
 (  -1.0,  -0.5)  &  29.247  &   22.194   &  17.135  &    9.381 &    12.297  &   13.144\\
 (  -0.5,   0.0)    & 6.892    &  5.178   &   3.945  &   -0.336   &   1.084  &    1.759\\
 (   0.0,   0.5)   &  6.857    &  5.152  &    3.927  &   -0.499  &    0.958  &   1.667\\
 (   0.5,   1.0)  & 29.224  & 22.182  & 17.133  &  9.631 &  12.455  &   13.243 \\ \hline \hline 
\multicolumn{7}{ c }{ $ 600~{\rm GeV} < \mttbar \le  800~{\rm GeV} $}  \\ \hline \hline 
 (  -1.0,  -0.5)  & 6.546  &  4.768 &   3.544 &   3.885 &    4.135     &  3.986\\
 (  -0.5,   0.0)  &  -2.559  &   -2.008  &  -1.606 & 0.231  &  -0.254 &    -0.501\\
 (   0.0,   0.5)  &  -2.561   &  -2.009   &  -1.606   &   0.199 &    -0.277&     -0.519\\
 (   0.5,   1.0)  &   6.575  &    4.787  &    3.554 &     3.972 &     4.198  &    4.043 \\ \hline \hline 
\multicolumn{7}{ c }{ $\mttbar > 800~{\rm GeV} $}  \\ \hline \hline 
 (  -1.0,  -0.5) &   -2.166  &   -1.628  &   -1.259 &     3.028  &    1.677 &     0.934\\
 (  -0.5,   0.0) &   -2.861   &  -2.130  &   -1.625  &    1.458  &    0.432 &    -0.097\\
 (   0.0,   0.5)  &  -2.861  &   -2.131  &   -1.627  &    1.489  &    0.447 &    -0.084\\
 (   0.5,   1.0) &   -2.173   &  -1.630   &  -1.257   &   2.985  &    1.647 &     0.899 \\ \hline \hline 
	\end{tabular} } 
	\end{table}%

		\begin{table}[h]
		\centering

		  {\renewcommand{\arraystretch}{1.2}
                  \renewcommand{\tabcolsep}{0.2cm}
		\caption{ \label{tab:binCrk} The numerator of the spin correlation $C(r,k)+C(k,r) = N_{rk}/\sigma_{\ttbar}$ in the SM
		at 13.6~TeV in bins of $\mttbar$ and $y_p=\cos\theta_t^*$ for 3 scales $\mu$.
		For each invariant mass bin the first column displays the range of the $y_p$ bin. The numbers in the 2nd, 3rd, and 4th column are 
		the values of $N_{0,rk}$ at LO QCD for $\mu=m_t/2, m_t,$ and $2m_t$, respectively. The 5th, 6th, and 7th column contain the
		NLO QCD plus weak-interaction  contributions $N_{1,rk}$ for $\mu=m_t/2, m_t,$ and $2m_t$, respectively.
		All numbers are in units of pb.}	
		\vspace{1mm}
\begin{tabular}{c c c c c c c }\hline	
\multicolumn{7}{ c }{ $ 2m_t\le \mttbar \le 450~{\rm GeV}$}  \\ \hline \hline   
 (  -1.0,  -0.5) & -17.107  & -13.884  & -11.423   & -6.623  & -7.214  &  -7.258 \\
 (  -0.5,   0.0)  & -10.935   & -8.831   & -7.231   & -3.927  &  -4.506  &  -4.653\\
 (   0.0,   0.5)  & -10.953    & -8.849   & -7.245   & -2.802   & -3.698  &  -4.062\\
 (   0.5,   1.0) & -17.092   & -13.866  & -11.410  &  -6.639    & -7.229  &  -7.277\\ \hline \hline 
\multicolumn{7}{ c }{ $ 450~{\rm GeV} < \mttbar\le 600~{\rm GeV} $}  \\ \hline \hline 
 (  -1.0,  -0.5)  & -20.499  & -16.052   & -12.784    & -5.027  &  -6.985  &  -7.670\\
 (  -0.5,   0.0) & -10.670   & -8.364   & -6.671    & 0.199    & -1.670  &  -2.602\\
 (   0.0,   0.5)  & -10.650   & -8.346   & -6.656   &  0.193  &  -1.661  &  -2.580\\
 (   0.5,   1.0)  & -20.505 &  -16.060   & -12.788  &  -5.131   & -7.058 &   -7.743\\ \hline \hline 
\multicolumn{7}{ c|}{ $ 600~{\rm GeV} < \mttbar \le  800~{\rm GeV} $}  \\ \hline \hline 
 (  -1.0,  -0.5)  & -10.152   & -7.723   & -5.994   & -0.561   & -2.286   & -3.009\\
 (  -0.5,   0.0)  & -3.923   & -3.000   & -2.343   &  1.136    & 0.065  &  -0.486\\
 (   0.0,   0.5)  & -3.927    & -3.004   & -2.343     & 1.046  &  -0.007  &  -0.541\\
 (   0.5,   1.0)  & -10.148   & -7.719   & -5.996   & -0.285   & -2.086    & -2.856\\ \hline \hline 
\multicolumn{7}{c}{ $\mttbar > 800~{\rm GeV} $}  \\ \hline \hline 
 (  -1.0,  -0.5)  & -4.592  &  -3.373   & -2.539    & 1.376   & -0.038   & -0.696\\
 (  -0.5,   0.0)  & -1.329    & -0.988  &  -0.752    & 0.868   &  0.319   &  0.034\\
 (   0.0,   0.5)   & -1.328    & -0.988   & -0.752    & 0.876   &  0.323  &   0.035\\
 (   0.5,   1.0)   & -4.600   & -3.377   & -2.542     & 1.357  &  -0.052  &  -0.706\\ \hline \hline 
\end{tabular}}
 
	\end{table}%

		\begin{table}[h]
		\centering
		 {\renewcommand{\arraystretch}{1.2}
                  \renewcommand{\tabcolsep}{0.2cm}
		\caption{ \label{tab:binCkks} The numerator of the spin correlation $C(k,k^*) = N_{kk^*}/\sigma_{\ttbar}$ in the SM
		at 13.6~TeV in bins of $\mttbar$ and $y=\cos\theta_t^*$ for 3 scales $\mu$.
		For each invariant mass bin the first column displays the range of the $y$ bin. The numbers in the 2nd, 3rd, and 4th column are 
		the values $N_{1,kk^*}$ at NLOW for $\mu=m_t/2, m_t,$ and $2m_t$, respectively.
		All numbers are in units of pb.}	
 \begin{tabular}{c c c c }\hline	
\multicolumn{4}{ c }{ $ 2m_t\le \mttbar \le 450~{\rm GeV}$}  \\ \hline \hline  	 
 (  -1.0,  -0.5) &  0.001  &    0.040  &    0.070  \\
 (  -0.5,   0.0) &    0.279 &     0.221 &     0.183\\
 (   0.0,   0.5) &   -0.304 &    -0.242 &    -0.192\\
 (   0.5,   1.0) &   -0.030 &    -0.019  &    0.003\\ \hline \hline 
\multicolumn{4}{ c }{ $ 450~{\rm GeV} < \mttbar\le 600~{\rm GeV} $}  \\ \hline \hline 
 (  -1.0,  -0.5)  &   0.366  &    0.314  &    0.278\\
 (  -0.5,   0.0)  &   0.008  &    0.016  &    0.009\\
 (   0.0,   0.5)  &   0.195 &     0.153 &     0.120\\
 (   0.5,   1.0)  &   0.613  &    0.520  &    0.453\\ \hline \hline 
\multicolumn{4}{ c }{ $ 600~{\rm GeV} < \mttbar \le  800~{\rm GeV} $}  \\ \hline \hline
 (  -1.0,  -0.5) &   -0.024  &    0.000 &     0.026\\
 (  -0.5,   0.0) &   -0.000 &     0.001  &    0.002\\
 (   0.0,   0.5) &    0.005   &   0.005  &    0.005\\
 (   0.5,   1.0) &   -0.037  &   -0.016  &    0.000\\ \hline \hline 
\multicolumn{4}{ c }{ $\mttbar > 800~{\rm GeV} $}  \\ \hline \hline 
 (  -1.0,  -0.5)  &   0.029  &    0.049  &    0.041\\
 (  -0.5,   0.0) &   -0.004  &   -0.002  &   -0.001\\
 (   0.0,   0.5) &    0.001 &     0.004 &     0.003\\
 (   0.5,   1.0)  &   0.069 &     0.073 &     0.071\\ \hline \hline 
\end{tabular} }
	\end{table}%

		\begin{table}[h]
		\centering
	
		 {\renewcommand{\arraystretch}{1.2}
                  \renewcommand{\tabcolsep}{0.2cm}
		\caption{ \label{tab:binCrks} The numerator of the spin correlation $C(r^*,k)+C(r,k^*) = N_{r^*k}/\sigma_{\ttbar}$  in the SM
		at 13.6~TeV in bins of $\mttbar$ and $y_p=\cos\theta_t^*$ for 3 scales $\mu$.
		For each invariant mass bin the first column displays the range of the $y_p$ bin. The numbers in the 2nd, 3rd, and 4th column are 
		the values $N_{1,r^*k}$ at NLOW for $\mu=m_t/2, m_t,$ and $2m_t$, respectively.
		All numbers are in units of pb.}		
		\vspace{1mm}

  \begin{tabular}{c c c c }\hline	
\multicolumn{4}{c}{ $ 2m_t\le \mttbar \le 450~{\rm GeV}$}  \\ \hline \hline  	
 (  -1.0,  -0.5) &  0.126  &    0.121 &     0.115 \\
 (  -0.5,   0.0) &   -0.161  &   -0.137 &    -0.108\\
 (   0.0,   0.5) &    0.042   &   0.037 &     0.031\\
 (   0.5,   1.0) &    0.041  &    0.052  &    0.066\\ \hline \hline  
\multicolumn{4}{c}{ $ 450~{\rm GeV} < \mttbar\le 600~{\rm GeV} $}  \\ \hline \hline 
 (  -1.0,  -0.5)  &   0.138  &    0.134  &    0.126\\
 (  -0.5,   0.0) &    0.059  &    0.046  &    0.035\\
 (   0.0,   0.5) &   -0.217 &    -0.168 &    -0.141\\
 (   0.5,   1.0) &   -0.127 &    -0.065  &   -0.031\\ \hline \hline  
\multicolumn{4}{c}{ $ 600~{\rm GeV} < \mttbar \le  800~{\rm GeV} $}  \\ \hline \hline 
 (  -1.0,  -0.5) &   -0.003  &    0.012  &    0.013\\
 (  -0.5,   0.0) &    0.006  &    0.006 &     0.002\\
 (   0.0,   0.5)  &   0.008   &   0.006  &    0.004\\
 (   0.5,   1.0)  &   0.042   &   0.027   &   0.029\\ \hline \hline  
\multicolumn{4}{c}{ $\mttbar > 800~{\rm GeV} $}  \\ \hline \hline 
 (  -1.0,  -0.5)  &   0.001  &   -0.001  &    0.009\\
 (  -0.5,   0.0) &   -0.002   &  -0.002  &   -0.001\\
 (   0.0,   0.5)  &   0.003   &   0.002  &    0.000\\
 (   0.5,   1.0) &   -0.000  &    0.004  &    0.005\\ \hline \hline  
 \end{tabular} }
	\end{table}%

		\begin{table}[h]
		\centering
		
		 {\renewcommand{\arraystretch}{1.2}
                  \renewcommand{\tabcolsep}{0.2cm}
		
		\caption{ \label{tab:binBr} The numerator of the polarization observable $B_1(r)+B_2(r) = N_{r}/\sigma_{\ttbar}$ 
		at 13.6~TeV in bins of $\mttbar$ and $y_p=\cos\theta_t^*$ for 3 scales $\mu$.
		For each invariant mass bin the first column displays the range of the $y_p$ bin. The numbers in the 2nd, 3rd, and 4th column are 
		the values  $N_{1,r}$ at NLOW for $\mu=m_t/2, m_t,$ and $2m_t$, respectively.
		All numbers are in units of pb.}		
		\vspace{1mm}
 \begin{tabular}{c c c c}\hline	
\multicolumn{4}{c}{ $ 2m_t\le \mttbar \le 450~{\rm GeV}$}  \\ \hline \hline  	
 (  -1.0,  -0.5) &  0.021 &     0.201  &    0.357 \\
 (  -0.5,   0.0) &    0.043 &     0.096  &    0.143\\
 (   0.0,   0.5)  &   0.043  &    0.096 &     0.143\\
 (   0.5,   1.0)  &   0.022  &    0.202  &    0.358\\ \hline \hline 
\multicolumn{4}{c}{ $ 450~{\rm GeV}<\mttbar\le 600~{\rm GeV} $}  \\ \hline \hline 
 (  -1.0,  -0.5) &    0.121  &    0.240  &    0.347\\
 (  -0.5,   0.0) &    0.110  &    0.113  &    0.119\\
 (   0.0,   0.5)  &   0.111 &     0.114  &    0.120\\
 (   0.5,   1.0) &    0.120 &     0.239  &    0.346\\ \hline \hline 
\multicolumn{4}{c}{ $ 600~{\rm GeV}< \mttbar \le  800~{\rm GeV} $}  \\ \hline \hline 
 (  -1.0,  -0.5)  &   0.096  &    0.116  &    0.139\\
 (  -0.5,   0.0) &    0.058   &   0.049 &     0.044\\
 (   0.0,   0.5) &    0.058  &    0.049  &    0.044\\
 (   0.5,   1.0) &    0.096  &    0.117  &    0.139\\ \hline \hline 
\multicolumn{4}{c}{ $\mttbar > 800~{\rm GeV} $}  \\ \hline \hline 
 (  -1.0,  -0.5)  &   0.061   &   0.056  &    0.056\\
 (  -0.5,   0.0)  &   0.025   &   0.019   &   0.015\\
 (   0.0,   0.5)  &   0.025  &    0.019 &     0.015\\
 (   0.5,   1.0)  &   0.061 &     0.056  &    0.056\\ \hline \hline 
 \end{tabular} }
	\end{table}%

		\begin{table}[h]
		\centering
	   {\renewcommand{\arraystretch}{1.2}
                  \renewcommand{\tabcolsep}{0.2cm}
		\caption{ \label{tab:binBk} The numerator of the polarization observable $B_1(k)+B_2(k) = N_{k}/\sigma_{\ttbar}$ 
		in the SM at 13.6~TeV in bins of $\mttbar$ and $y_p=\cos\theta_t^*$ for 3 scales $\mu$.
		For each invariant mass bin the first column displays the range of the $y_p$ bin. The numbers in the 2nd, 3rd, and 4th column are 
		the values of $N_{1,k}$ at NLOW for $\mu=m_t/2, m_t,$ and $2m_t$, respectively.
		All numbers are in units of pb.}		
		\vspace{1mm}
 \begin{tabular}{c c c c}\hline	
\multicolumn{4}{c}{ $ 2m_t\le \mttbar \le 450~{\rm GeV}$}  \\ \hline \hline  	
 (  -1.0,  -0.5) &  0.058   & 0.285  &  0.482\\
 (  -0.5,   0.0)  & 0.047 &   0.007  & -0.025\\
 (   0.0,   0.5)  & 0.046  &  0.007  & -0.026\\
 (   0.5,   1.0) &  0.061  &  0.287  &  0.485\\ \hline \hline  
\multicolumn{4}{c}{ $ 450~{\rm GeV} < \mttbar\le 600~{\rm GeV} $}  \\ \hline \hline 
 (  -1.0,  -0.5) &  0.428   & 0.630   & 0.817\\
 (  -0.5,   0.0) &  0.181 &   0.114  &  0.063\\
 (   0.0,   0.5) &  0.181 &   0.114  &  0.063\\
 (   0.5,   1.0)  & 0.426  &  0.628  &  0.815\\ \hline \hline  
\multicolumn{4}{c}{ $ 600~{\rm GeV} < \mttbar \le  800~{\rm GeV} $}  \\ \hline \hline 
 (  -1.0,  -0.5)  & 0.484  &  0.531  &  0.587\\
 (  -0.5,   0.0) &  0.150  &  0.104 &   0.071\\
 (   0.0,   0.5) &  0.150  &  0.104  &  0.071\\
 (   0.5,   1.0) &  0.485  &  0.531  &  0.588\\ \hline \hline  
\multicolumn{4}{c}{ $\mttbar > 800~{\rm GeV} $}  \\ \hline \hline 
 (  -1.0,  -0.5) &  0.531  &  0.494  &  0.485\\
 (  -0.5,   0.0) &  0.121  &  0.085  &  0.060\\
 (   0.0,   0.5)  & 0.121   & 0.085   & 0.060\\
 (   0.5,   1.0)  & 0.532  &  0.494  &  0.485\\ \hline \hline  
 	\end{tabular} }
	\end{table}%

		\begin{table}[h]
		\centering
		{\renewcommand{\arraystretch}{1.2}
                  \renewcommand{\tabcolsep}{0.2cm}
		\caption{ \label{tab:binBrs} The numerator of the polarization observable $B_1(r^*)+B_2(r^*) = N_{r^*}/\sigma_{\ttbar}$ 
		in the SM at 13.6~TeV in bins of $\mttbar$ and $y_p=\cos\theta_t^*$ for 3 scales $\mu$.
		For each invariant mass bin the first column displays the range of the $y_p$ bin. The numbers in the 2nd, 3rd, and 4th column are 
		the values of $N_{1,r^*}$ at NLOW for $\mu=m_t/2, m_t,$ and $2m_t$, respectively.
		All numbers are in units of pb.}
		\vspace{1mm}
  \begin{tabular}{c c c c}\hline	
\multicolumn{4}{c}{ $ 2m_t\le \mttbar \le 450~{\rm GeV}$}  \\ \hline \hline  
 (  -1.0,  -0.5) & -0.036   &  0.013   &  0.054\\ 
 (  -0.5,   0.0)  & -0.037    & 0.013    & 0.053\\ 
 (   0.0,   0.5)   & -0.036   &  0.013   &  0.054\\ 
 (   0.5,   1.0)  & -0.035   &  0.014    & 0.055\\ \hline \hline  
\multicolumn{4}{c}{ $ 450~{\rm GeV} < \mttbar\le 600~{\rm GeV} $}  \\ \hline \hline 
 (  -1.0,  -0.5)  & -0.030   &  0.002   &  0.028\\ 
 (  -0.5,   0.0)  & -0.009    & 0.007   &  0.020\\ 
 (   0.0,   0.5)  & -0.009     & 0.007   &  0.020\\ 
 (   0.5,   1.0)  & -0.031    & 0.001   &  0.027\\ \hline \hline  
\multicolumn{4}{c}{ $ 600~{\rm GeV} < \mttbar \le  800~{\rm GeV} $}  \\ \hline \hline 
 (  -1.0,  -0.5)  & -0.013   & -0.003  &   0.005\\ 
 (  -0.5,   0.0)  &  0.007  &   0.008  &   0.009\\ 
 (   0.0,   0.5)   & 0.006    & 0.008   &  0.009\\ 
 (   0.5,   1.0)   & -0.013  &  -0.003   &  0.005\\ \hline \hline  
\multicolumn{4}{c}{ $\mttbar > 800~{\rm GeV} $}  \\ \hline \hline 
 (  -1.0,  -0.5)  & -0.006   & -0.003   & 0.000\\ 
 (  -0.5,   0.0)   & 0.009    & 0.008   &  0.006\\ 
 (   0.0,   0.5)   & 0.009    & 0.008   &  0.007\\ 
 (   0.5,   1.0)   & -0.006   & -0.003   &  0.000\\ \hline \hline  
 \end{tabular}}
	\end{table}%

		\begin{table}[h]
		\centering
		{\renewcommand{\arraystretch}{1.2}
                  \renewcommand{\tabcolsep}{0.2cm}
		\caption{ \label{tab:binBks} The numerator of the polarization observable $B_1(k^*)+B_2(k^*) = N_{k^*}/\sigma_{\ttbar}$ 
		in the SM at 13.6~TeV in bins of $\mttbar$ and $y_p=\cos\theta_t^*$ for 3 scales $\mu$.
		For each invariant mass bin the first column displays the range of the $y_p$ bin. The numbers in the 2nd, 3rd, and 4th column are 
		the values of $N_{1,k^*}$ at NLOW for $\mu=m_t/2, m_t,$ and $2m_t$, respectively.
		All numbers are in units of pb.}		
		\vspace{1mm}
  \begin{tabular}{c c c c }\hline	
\multicolumn{4}{c}{ $ 2m_t\le \mttbar \le 450~{\rm GeV}$}  \\ \hline \hline  	
 (  -1.0,  -0.5) &  -0.040  &   0.028  &   0.083\\ 
 (  -0.5,   0.0)  & -0.008  &   0.005  &   0.015\\ 
 (   0.0,   0.5) &  -0.008  &   0.005  &   0.015\\ 
 (   0.5,   1.0) &  -0.038  &   0.030  &   0.086\\ \hline \hline  	
\multicolumn{4}{c}{ $ 450~{\rm GeV} < \mttbar\le 600~{\rm GeV} $}  \\ \hline \hline 
 (  -1.0,  -0.5) &  -0.023  &   0.037  &   0.086\\ 
 (  -0.5,   0.0) &   0.001  &   0.005  &   0.008\\ 
 (   0.0,   0.5) &   0.001 &    0.005  &   0.008\\ 
 (   0.5,   1.0) &  -0.025  &   0.035  &   0.084\\ \hline \hline  	
\multicolumn{4}{c}{ $ 600~{\rm GeV} < \mttbar \le  800~{\rm GeV} $}  \\ \hline \hline 
 (  -1.0,  -0.5) &  -0.006  &   0.023  &   0.047\\ 
 (  -0.5,   0.0) &   0.006  &  0.005  &  0.005\\ 
 (   0.0,   0.5) &  0.006  &  0.005 &   0.005\\ 
 (   0.5,   1.0)&  -0.005  &  0.024  &  0.048\\ \hline \hline  	
\multicolumn{4}{c}{ $\mttbar > 800~{\rm GeV} $}  \\ \hline \hline 
 (  -1.0,  -0.5)  & 0.005   & 0.021  &  0.035\\ 
 (  -0.5,   0.0)  & 0.009  &  0.007 &   0.005\\ 
 (   0.0,   0.5) &  0.009   & 0.007 &   0.005\\ 
 (   0.5,   1.0)  & 0.005   & 0.021  &  0.035\\ \hline \hline  	
 \end{tabular}}
	\end{table}%

			\begin{table}[h]
		\centering
		{\renewcommand{\arraystretch}{1.2}
                  \renewcommand{\tabcolsep}{0.2cm}
		\caption{ \label{tab:binBabs} The numerator of the polarization observable $B_1(n)+B_2(n) = N_{n}/\sigma_{\ttbar}$ 
		in the SM at 13.6~TeV in bins of $\mttbar$ and $y_p=\cos\theta_t^*$ for 3 scales $\mu$.
		For each invariant mass bin the first column displays the range of the $y_p$ bin. The numbers in the 2nd, 3rd, and 4th column are 
		the values $N_{1,n}$ at NLO QCD for $\mu=m_t/2, m_t,$ and $2m_t$, respectively.
		All numbers are in units of pb.}		
		\vspace{1mm}
 \begin{tabular}{c c c c}\hline	
\multicolumn{4}{c}{ $ 2m_t\le \mttbar \le 450~{\rm GeV}$}  \\ \hline \hline  	
 (  -1.0,  -0.5) &0.260  & 0.187    &  0.138 \\
 (  -0.5,   0.0) & 0.184 &  0.132  &  0.097\\
 (   0.0,   0.5) & 0.184  & 0.132  & 0.097 \\
 (   0.5,   1.0) &  0.260 & 0.187  & 0.138    \\ \hline \hline  	
\multicolumn{4}{c}{ $ 450~{\rm GeV} < \mttbar\le 600~{\rm GeV} $}  \\ \hline \hline 
 (  -1.0,  -0.5) & 0.761  & 0.533   & 0.384 \\
 (  -0.5,   0.0) &  0.457   & 0.320  &  0.230    \\
 (   0.0,   0.5) &  0.457  & 0.320  &  0.230 \\
 (   0.5,   1.0) &  0.761 &  0.533 &  0.384    \\ \hline \hline  	
\multicolumn{4}{c}{ $ 600~{\rm GeV} < \mttbar \le  800~{\rm GeV} $}  \\ \hline \hline 
 (  -1.0,  -0.5) & 0.546  & 0.373  &  0.263 \\
 (  -0.5,   0.0) &  0.261   & 0.179 &  0.126 \\
 (   0.0,   0.5)  & 0.261  &  0.179 &  0.126    \\
 (   0.5,   1.0) & 0.546  & 0.373  &  0.263 \\ \hline \hline  	
\multicolumn{4}{c}{ $\mttbar > 800~{\rm GeV} $}  \\ \hline \hline 
 (  -1.0,  -0.5) & 0.310  & 0.204   &  0.140 \\
 (  -0.5,   0.0)  & 0.113   & 0.075   & 0.052  \\
 (   0.0,   0.5)  & 0.113 &  0.075  &   0.052 \\
 (   0.5,   1.0)   & 0.310  & 0.204  & 0.140  	\\ \hline \hline  		
		\end{tabular} }
	\end{table}%

			\begin{table}[h]
		\centering
		{\renewcommand{\arraystretch}{0.95}
                  \renewcommand{\tabcolsep}{0.2cm}
		
		\caption{ \label{tab:binNP1hm} 
			The  NP contributions to the $\ttbar$ cross section,
			to the numerator of $A_C$, and to the numerator of $C(n,n)$
		at 13.6~TeV in bins of $\mttbar$ and $y_p=\cos\theta_t^*$ for the scale  $\mu=m_t/2.$
		For each invariant mass bin listed in the first column the second to  fifth column 
		 shows the respective contribution in the indicated range of the $y_p$ bin. 
		All numbers are in units of pb.}
		\vspace{1mm}
 \begin{tabular}{l c c c c }\hline		
          &(  -1.0,  -0.5) & (  -0.5,   0.0) &	(   0.0,   0.5) &(   0.5,   1.0) \\ \hline \hline  	
\multicolumn{5}{c}{ NP contribution $\propto \hcvv$  to $\stt$.  }  \\ \hline  			
$(2m_t,450~{\rm GeV})$  & 52.055     &    48.679 &        48.678&        52.058            \\
$(450~{\rm GeV}, 600~{\rm GeV})$   &  62.353&        52.787 &        52.793 &        62.353  \\     
$(600~{\rm GeV}, 800~{\rm GeV})$ &  46.484 &       36.325 &        36.328 &       46.480  \\    
 $ > 800 ~{\rm GeV}$ &   66.005 &       47.819 &        47.818 &        66.006   \\ \hline   
 \multicolumn{5}{c}{ NP contribution $\propto \xx1$  to $\stt$.  }  \\ \hline  
 $(2m_t,450~{\rm GeV})$         &  5.497 &      5.138 &        5.139 &       5.497 \\
\hline \hline      
 		\end{tabular} }
	\end{table}%

	\clearpage

			\begin{table}[t]
		\centering
		{\renewcommand{\arraystretch}{0.95}
                  \renewcommand{\tabcolsep}{0.2cm}
		
		\caption{ \label{tab:binNP2hm} 
		Continuation of Table~\ref{tab:binNP1hm}. The  NP contributions to the binned numerators
		 of the displayed spin observables  at 13.6~TeV for the scale  $\mu=m_t/2.$
		All numbers are in units of pb.}
		\vspace{1mm}
 \begin{tabular}{l c c c c }\hline		
          &(  -1.0,  -0.5) & (  -0.5,   0.0) &	(   0.0,   0.5) &(   0.5,   1.0) \\ \hline \hline  	
   \multicolumn{5}{c}{ NP contribution $\propto \hcvv$  to numerator of $C(r,r)$.  }  \\ \hline   
   $(2m_t,450~{\rm GeV})$  &  -20.041 &       -44.092 &      -44.092 &       -20.042   \\    
    $(450~{\rm GeV}, 600~{\rm GeV})$ & -21.322 &    -46.909 &      -46.912 &      -21.321 \\    
  $(600~{\rm GeV}, 800~{\rm GeV})$ & -14.424 &       -31.733 &      -31.730 &      -14.424 \\     
 $ > 800 ~{\rm GeV}$ &  -18.650 &      -41.031 &      -41.028 &      -18.650 \\   \hline  
     \multicolumn{5}{c}{ NP contribution $\propto \xx1$  to numerator of $C(r,r)$.  }  \\ \hline   
    $(2m_t,450~{\rm GeV})$  &  -2.115 &       -4.654 &       -4.654 &       -2.115  \\
\hline   
 		\end{tabular} }
	\end{table}%

			\begin{table}[t]
		\centering
		{\renewcommand{\arraystretch}{0.95}
                  \renewcommand{\tabcolsep}{0.2cm}
		
		\caption{ \label{tab:binNP3hm} 
		Continuation of Table~\ref{tab:binNP2hm}. 
		 The  NP contributions to the binned numerators
		 of the displayed spin observables  at 13.6~TeV for the scale  $\mu=m_t/2.$
		All numbers are in units of pb.}
		\vspace{1mm}
 \begin{tabular}{l c c c c}\hline		
          &(  -1.0,  -0.5) & (  -0.5,   0.0) &	(   0.0,   0.5) &(   0.5,   1.0) \\ \hline \hline 	
        \multicolumn{5}{c}{ NP contribution $\propto \Hmu$  to numerator of $C(r,k)+C(k,r)$.  } \\ \hline  
       $(2m_t,450~{\rm GeV})$  &  -3.809 &      -20.567&       -20.567 &      -3.809  \\     
    $(450~{\rm GeV}, 600~{\rm GeV})$ &  2.673 &       -30.307 &      -30.308 &       2.676 \\    
    $(600~{\rm GeV}, 800~{\rm GeV})$ & -22.284 &       -23.897 &      -23.899 &      -22.287    \\     
   $ > 800 ~{\rm GeV}$ &  -48.061 &      -20.339 &      -20.339    &    -48.060 \\  \hline   \hline 
  \multicolumn{5}{c}{ NP contribution $\propto \Hd$  to numerator of $C(n,r)-C(r,n)$.  }  \\ \hline  
   $(2m_t,450~{\rm GeV})$  & -261.754 &       -297.544 &       -297.549 &       -261.750    \\
\hline   \hline  	
 		\end{tabular} }
	\end{table}%

			\begin{table}[t]
		\centering
		{\renewcommand{\arraystretch}{0.95}
                  \renewcommand{\tabcolsep}{0.2cm}
		
		\caption{ \label{tab:binNP4hm} 
		Continuation of Table~\ref{tab:binNP3hm}. 
				 The  NP contributions to the binned numerators
		 of the displayed spin observables  at 13.6~TeV for the scale  $\mu=m_t/2.$
		All numbers are in units of pb.}
		\vspace{1mm}
 \begin{tabular}{l c c c c}\hline		
          &(  -1.0,  -0.5) & (  -0.5,   0.0) &	(   0.0,   0.5) &(   0.5,   1.0) \\ \hline \hline 	
        \multicolumn{5}{c}{ NP contribution $\propto \hcva$  to numerator of $B_1(r)+B_2(r)$.}  \\ \hline  	
  $(2m_t,450~{\rm GeV})$  &  9.855 &        5.319 &       5.319 &        9.855   \\    
  $(450~{\rm GeV}, 600~{\rm GeV})$ &    15.032 &       8.113 &        8.113 &      15.032   \\     
  $(600~{\rm GeV}, 800~{\rm GeV})$ &  10.349 &       5.585 &        5.585 &        10.349 \\    
  $ > 800 ~{\rm GeV}$ &   10.059 &       5.429 &        5.429 &        10.059  \\   \hline   
        \multicolumn{5}{c}{ NP contribution $\propto \xz3$  to numerator of $B_1(r)+B_2(r)$.}  \\ \hline   
  $(2m_t,450~{\rm GeV})$  & 1.044 &      0.563 &       0.563 &        1.044  \\
\hline   \hline     
 		\end{tabular} }
	\end{table}%
	
\clearpage	
		\begin{table}[t]
		\centering
		{\renewcommand{\arraystretch}{0.95}
                  \renewcommand{\tabcolsep}{0.2cm}
		
		\caption{ \label{tab:binNP1mm} 
		As in table~\ref{tab:binNP1hm},  but for the scale  $\mu=m_t.$
		 All numbers are in units of pb.}
	\vspace{1mm}	
 \begin{tabular}{l c c c c}\hline		
          &(  -1.0,  -0.5) & (  -0.5,   0.0) &	(   0.0,   0.5) &(   0.5,   1.0) \\ \hline \hline  	
\multicolumn{5}{c}{ NP contribution $\propto \hcvv$  to $\stt$.  }  \\ \hline  
 $(2m_t,450~{\rm GeV})$  &  43.315  &   40.512     &   40.511    &   43.315  \\     
 $(450~{\rm GeV}, 600~{\rm GeV})$ &   51.091   &   43.258   &    43.261    &    51.089    \\    
  $(600~{\rm GeV}, 800~{\rm GeV})$ &    37.408  &    29.235 &    29.238   &     37.405   \\    
  $ > 800 ~{\rm GeV}$ &    51.072 &       37.023  &    37.023 &    51.074 \\  \hline     
\multicolumn{5}{c}{ NP contribution $\propto \xx1$  to $\stt$.  }  \\ \hline    
$(2m_t,450~{\rm GeV})$   &  4.643 &   4.340    &  4.341 &     4.643   \\
\hline  \hline   
 		\end{tabular} }
	\end{table}%
  
\clearpage	
	
		\begin{table}[t]
		\centering
		{\renewcommand{\arraystretch}{0.95}
                  \renewcommand{\tabcolsep}{0.2cm}
	
		\caption{ \label{tab:binNP2mm} As in table~\ref{tab:binNP2hm},  but for the scale  $\mu=m_t.$
		 All numbers are in units of pb.}
		\vspace{1mm}
 \begin{tabular}{l c c c c }\hline		
          &(  -1.0,  -0.5) & (  -0.5,   0.0) &	(   0.0,   0.5) &(   0.5,   1.0) \\ \hline \hline  	
 \multicolumn{5}{c}{ NP contribution $\propto \hcvv$  to numerator of $C(r,r)$.  }  \\ \hline   	
     $(2m_t,450~{\rm GeV})$ &   -16.679 & -36.694  & -36.694 &   -16.679 \\   
      $(450~{\rm GeV}, 600~{\rm GeV})$ &   -17.473 & -38.441 & -38.439  &    -17.473  \\     
    $(600~{\rm GeV}, 800~{\rm GeV})$ &  -11.609  & -25.539 &  -25.536 &   -11.608 \\    
      $ > 800 ~{\rm GeV}$ &  -14.442 &   -31.772    &  -31.771 &     -14.442  \\ \hline   
   \multicolumn{5}{c}{ NP contribution $\propto \xx1$  to numerator of $C(r,r)$.  }  \\ \hline     
 $(2m_t,450~{\rm GeV})$ &   -1.787  &  -3.931 &   -3.931 &    -1.787 \\
\hline  \hline  
 		\end{tabular} }
	\end{table}%
 
\clearpage	
	
		\begin{table}[t]
		\centering
		{\renewcommand{\arraystretch}{0.95}
                  \renewcommand{\tabcolsep}{0.2cm}
		\caption{ \label{tab:binNP3mm} As in table~\ref{tab:binNP3hm},  but for the scale  $\mu=m_t.$
		 All numbers are in units of pb. }
		\vspace{1mm}
 \begin{tabular}{l c c c c }\hline		
          &(  -1.0,  -0.5) & (  -0.5,   0.0) &	(   0.0,   0.5) &(   0.5,   1.0) \\ \hline \hline  
 \multicolumn{5}{c}{ NP contribution $\propto \Hmu$  to numerator of $C(r,k)+C(k,r)$.  }  \\ \hline          
   $(2m_t,450~{\rm GeV})$ &   -5.148   &   -17.288 & -17.289 &   -5.147   \\    
     $(450~{\rm GeV}, 600~{\rm GeV})$ & 0.326 &  -24.253  &  -24.253 &   0.328 \\    
       $(600~{\rm GeV}, 800~{\rm GeV})$ & -17.617 &     -18.435 &  -18.436  &   -17.616\\    
     $ > 800 ~{\rm GeV}$ &   -35.158  &   -15.070 &  -15.070 &  -35.158 \\ \hline  \hline   
  \multicolumn{5}{c}{ NP contribution $\propto \Hd$  to numerator of $C(n,r)-C(r,n)$.  }  \\ \hline    
    $(2m_t,450~{\rm GeV})$ &  -204.974 &   -232.705 & -232.704 &   -204.972 \\
\hline  \hline  	
 		\end{tabular} }
	\end{table}%
 
\clearpage
	
		\begin{table}[t]
		\centering
		{\renewcommand{\arraystretch}{0.95}
                  \renewcommand{\tabcolsep}{0.2cm}
		\caption{ \label{tab:binNP4mm} As in table~\ref{tab:binNP4hm},  but for the scale  $\mu=m_t.$
		 All numbers are in units of pb.}
		\vspace{1mm}
 \begin{tabular}{l c c c c}\hline		
          &(  -1.0,  -0.5) & (  -0.5,   0.0) &	(   0.0,   0.5) &(   0.5,   1.0) \\ \hline \hline  
       \multicolumn{5}{c}{ NP contribution $\propto \hcva$  to numerator of $B_1(r)+B_2(r)$.}  \\ \hline  
 $(2m_t,450~{\rm GeV})$ &    8.193 &  4.422 &   4.422 &  8.193 \\     
 $(450~{\rm GeV}, 600~{\rm GeV})$ &   12.316  &  6.647 &   6.647&  12.317  \\    
 $(600~{\rm GeV}, 800~{\rm GeV})$ &  8.330   &  4.496 &  4.495 &    8.330\\     
  $ > 800 ~{\rm GeV}$ & 7.832 &   4.227  &   4.227  &    7.832 \\  \hline     
     \multicolumn{5}{c}{ NP contribution $\propto \xz3$  to numerator of $B_1(r)+B_2(r)$.}  \\ \hline   
  $(2m_t,450~{\rm GeV})$ &  0.881    &  0.475 &   0.475 &  0.881 \\
\hline   \hline  
 		\end{tabular} }
	\end{table}%
 \clearpage	
		\begin{table}[t]
		\centering
		{\renewcommand{\arraystretch}{0.95}
                  \renewcommand{\tabcolsep}{0.2cm}
		\caption{ \label{tab:binNP12m} 
	As in table~\ref{tab:binNP1hm},  but for the scale  $\mu=2 m_t.$
		 All numbers are in units of pb.}
		\vspace{1mm}
 \begin{tabular}{l c c c c}\hline		
          &(  -1.0,  -0.5) & (  -0.5,   0.0) &	(   0.0,   0.5) &(   0.5,   1.0) \\ \hline \hline  	
\multicolumn{5}{c}{ NP contribution $\propto \hcvv$  to $\stt$.  }  \\ \hline  
     $(2m_t,450~{\rm GeV})$ &    36.475 &   34.116      &    34.119  &   36.475    \\     
     $(450~{\rm GeV}, 600~{\rm GeV})$ & 42.441   &  35.938 &  35.942  &    42.442   \\    
   $(600~{\rm GeV}, 800~{\rm GeV})$ &   30.584 &   23.905   &  23.908        & 30.582 \\    
   $ > 800 ~{\rm GeV}$ &   40.357 &  29.271 &     29.270  &  40.357\\ \hline      
 \multicolumn{5}{c}{ NP contribution $\propto \xx1$  to $\stt$.  }  \\ \hline   
    $(2m_t,450~{\rm GeV})$ &   3.961  & 3.703 &  3.703 &   3.960    \\
\hline    \hline    
 		\end{tabular} }
	\end{table}%
\clearpage	
		\begin{table}[t]
		\centering
		{\renewcommand{\arraystretch}{0.95}
                  \renewcommand{\tabcolsep}{0.2cm}
		\caption{ \label{tab:binNP22m} As in table~\ref{tab:binNP2hm},  but for the scale  $\mu=2 m_t.$
		 All numbers are in units of pb.}
		\vspace{1mm}
 \begin{tabular}{l c c c c}\hline		
          &(  -1.0,  -0.5) & (  -0.5,   0.0) &	(   0.0,   0.5) &(   0.5,   1.0) \\ \hline \hline  
  \multicolumn{5}{c}{ NP contribution $\propto \hcvv$  to numerator of $C(r,r)$.  }  \\ \hline          
      $(2m_t,450~{\rm GeV})$ & -14.047  &  -30.903 &   -30.904 &     -14.047 \\     
     $(450~{\rm GeV}, 600~{\rm GeV})$ &  -14.517 & -31.937 &   -31.939 &   -14.517     \\    
  $(600~{\rm GeV}, 800~{\rm GeV})$ &   -9.493 &   -20.884     & -20.882 &  -9.492  \\     
   $ > 800 ~{\rm GeV}$ & -11.419 &    -25.121  &   -25.122 &  -11.419\\   \hline       
  \multicolumn{5}{c}{ NP contribution $\propto \xx1$  to numerator of $C(r,r)$.  }  \\ \hline    
      $(2m_t,450~{\rm GeV})$ & -1.524  &   -3.354    &  -3.354     &  -1.524     \\
\hline\hline 
 		\end{tabular} }
	\end{table}%
\clearpage	
		\begin{table}[t]
		\centering
		{\renewcommand{\arraystretch}{0.95}
                  \renewcommand{\tabcolsep}{0.2cm}
		\caption{ \label{tab:binNP32m} As in table~\ref{tab:binNP3hm},  but for the scale  $\mu=2 m_t.$
		 All numbers are in units of pb.}
		\vspace{1mm}
 \begin{tabular}{l c c c c}\hline		
          &(  -1.0,  -0.5) & (  -0.5,   0.0) &	(   0.0,   0.5) &(   0.5,   1.0) \\ \hline \hline  	
  \multicolumn{5}{c}{ NP contribution $\propto \Hmu$  to numerator of $C(r,k)+C(k,r)$.  }  \\ \hline             
       $(2m_t,450~{\rm GeV})$ &  -5.820 &     -14.686 &       -14.687 &       -5.819\\    
     $(450~{\rm GeV}, 600~{\rm GeV})$ &  -1.104 &      -19.720 &      -19.720 &     -1.104 \\    
   $(600~{\rm GeV}, 800~{\rm GeV})$ &  -14.182 &   -14.508    &   -14.507 &  -14.181   \\     
    $ > 800 ~{\rm GeV}$ &    -26.424 &     -11.453  &   -11.453 &       -26.424 \\ \hline  \hline     
  \multicolumn{5}{c}{ NP contribution $\propto \Hd$  to numerator of $C(n,r)-C(r,n)$.  }  \\ \hline    
       $(2m_t,450~{\rm GeV})$ &  -162.885 &  -184.656 &   -184.655 &  -162.877 \\
\hline  \hline   
 		\end{tabular} }
	\end{table}%
\clearpage	
		\begin{table}[t]
		\centering
		{\renewcommand{\arraystretch}{0.95}
                  \renewcommand{\tabcolsep}{0.2cm}
		\caption{ \label{tab:binNP42m} As in table~\ref{tab:binNP4hm},  but for the scale  $\mu=2 m_t.$
		 All numbers are in units of pb.}
		\vspace{1mm}
 \begin{tabular}{l c c c c }\hline		
          &(  -1.0,  -0.5) & (  -0.5,   0.0) &	(   0.0,   0.5) &(   0.5,   1.0) \\ \hline \hline   
       \multicolumn{5}{c}{ NP contribution $\propto \hcva$  to numerator of $B_1(r)+B_2(r)$.}  \\ \hline            
  $(2m_t,450~{\rm GeV})$ &  6.894 &       3.721 &        3.721 &        6.894 \\     
     $(450~{\rm GeV}, 600~{\rm GeV})$ &  10.231 &    5.522 &    5.522   &   10.231 \\     
    $(600~{\rm GeV}, 800~{\rm GeV})$ &   6.812 &   3.676 &   3.676     &    6.812   \\    
    $ > 800 ~{\rm GeV}$ &    6.222 &   3.358  &       3.358 &  6.222 \\    \hline  
        \multicolumn{5}{c}{ NP contribution $\propto \xz3$  to numerator of $B_1(r)+B_2(r)$.}  \\ \hline  
   $(2m_t,450~{\rm GeV})$ & 0.751 &       0.405 &      0.405 &      0.751 \\
\hline    \hline   
 		\end{tabular} }
	\end{table}%
\clearpage	
		

\pagebreak	
\newpage	
	

\end{document}